\documentclass[12pt]{article}

\usepackage{newtxtext,newtxmath}

\usepackage{graphicx}
\usepackage{color}
\usepackage[letterpaper,margin=1in]{geometry}

\renewenvironment{abstract}
	{\quotation}
	{\endquotation}

\date{}

\makeatletter
\renewcommand{\fnum@figure}{\textbf{Figure \thefigure}}
\renewcommand{\fnum@table}{\textbf{Table \thetable}}
\makeatother

\usepackage{scicite}
\usepackage{url}

\def\scititle{
	Unidirectional exceptional point of reflectionless states in a magnonic mirror array
}
\title{\bfseries \boldmath \scititle}

\author{
	Zi-Qi Wang$^{1\ast}$,
	Yuan-Peng Peng$^{1\ast}$,
    Yi-Pu Wang$^{1\dagger}$,
    J. Q. You$^{1,2\ddagger}$\and
	\small$^{1}$Zhejiang Key Laboratory of Micro-Nano Quantum Chips and Quantum Control, \and \small School of Physics, and State Key Laboratory for Extreme Photonics and Instrumentation, \and \small Zhejiang University, Hangzhou 310027, China.\\[1mm]
    \small$^{2}$College of Optical Science and Engineering, Zhejiang University, Hangzhou 310027, China\\[1mm]
	\small$^\ast$ These authors contributed equally to this work. \\[-3mm]
	\small$^\dagger$ Corresponding author. Email: yipuwang@zju.edu.cn \\[-3mm]
	\small$^\ddagger$ Corresponding author. Email: jqyou@zju.edu.cn 		
}

\begin{document} 

\maketitle

\begin{abstract} \bfseries \boldmath
Exceptional points (EPs) in non-Hermitian systems are singularities where both eigenvalues and eigenvectors coalesce. In scattering systems, EPs correspond to the merging of scattering states, leading to reflectionless (RL) behavior. A reflectionless exceptional point (RL EP) arises when two RL states further coalesce, yielding an anomalous quartic spectral response. While RL EPs have been explored in bidirectional systems, their unidirectional realization remains elusive. Here, we experimentally demonstrate a unidirectional RL EP by engineering collective states in an anti-Bragg magnonic mirror array. Inversion symmetry is broken using a giant spin ensemble that couples to a waveguide at three spatially separated points, enabling unidirectional reflectionless. At the RL EP, the reflection spectrum flattens and broadens significantly beyond the Lorentzian profile. The observed spectral valleys also expose dark-state behaviors that are typically inaccessible through conventional measurements. Our results provide a route toward controlling collective coherence in open systems and developing broadband unidirectional devices.
\end{abstract}
\paragraph{Short title:} Unidirectional reflectionless magnonic mirror array at an exceptional point
\paragraph{Teaser:}At the exceptional point, a unidirectional reflectionless magnonic mirror array widens the silence of reflection into a broad band.
\noindent
\subsection*{Introduction}
Exceptional points (EPs) in non-Hermitian systems correspond to degeneracies where both eigenvalues and eigenvectors coalesce, often giving rise to anomalous wave responses~\cite{Feng-13,Lin-11,longhi-11,Jones-12,Soleymani-22}. In scattering systems, scattering EPs are typically associated with the formation of reflectionless (RL) states, where coalescence of scattering eigenchannels yields zero reflection at a particular frequency. Unidirectional RL states, in which reflection vanishes in one direction but remains strong in the opposite, are especially appealing for applications in unidirectional invisibility. This concept has been extensively investigated in parity-time (PT) symmetric optical structures~\cite{Feng-13,Feng-14,Lin-11,longhi-11,Huang-17,Yin-13,Jia-15,Jones-12}, and can be adapted to realize asymmetric reflection between distinct polarization states in non‑Hermitian metasurfaces~\cite{Qin-25,Lawrence-14,Yang-24}. Nevertheless, the unidirectional RL states induced by EPs in previous studies are generally narrowband~\cite{Wu-14,Qian-23,Rao-21,Han-23,AnZ-24,Qin-25,Lawrence-14,Yang-24,Feng-14,Lin-11} limited by the Lorentzian profile of isolated resonances. Recently, a special type of EP, known as the RL EP, has been proposed to enhance spectral performance~\cite{Ferise-22,Stone-19,Stone-20,Yang-21,Rotter-24}. At an RL EP, two degenerate reflectionless solutions merge, giving rise to a quartic spectral response that significantly broadens the reflectionless bandwidth without requiring external gain or complex control parameters~\cite{Yang-17,Hodaei-17,AluL-19,Wiersig-20,Vahala-20,Kononchuk-22,Wiersig-14,AnZ-24}. Up to now, all RL EPs demonstrated have been confined to symmetric or bidirectional platforms. The emergence of unidirectional behavior requires an additional ingredient involving the breaking of spatial symmetry such as inversion or mirror symmetry in the system's geometry or dissipation. Thus, experimentally achieving and controling the necessary asymmetry within a fabricated sample remains a significant challenge.

A radiative emitter array coupled to the waveguide serves as a promising experimental platform for studying these unusual scattering phenomena~\cite{Roy-17,Sheremet-23,VanLoo-13,corzo-19,Mukhopadhyay-19,Ziqi-22}. Each emitter in such a system can act as a mirror at resonance and its effective reflectivity is determined by its radiation efficiency. Notably, employing a giant emitter beyond the dipole approximation allows for flexible control of reflectivity, as its radiative damping can be modulated through the resonance frequency~\cite{Kockum-14,Kockum-18,Guo-17A,Cirac-19,Kannan-20,Andersson-19,Kockum-21,Wang-21L,peng-23,Vadiraj-21A,Roccati-24,Du-22}. Moreover, the addition of coupling points can further augments this capability, as the collective interference in the giant emitter can significantly enhance its radiation efficiency~\cite{Vadiraj-21A}. The unique tunability of the giant emitter allows for both the generation and precise adjustment of system asymmetry, facilitating the emergence of novel collective effects. In contrast to the conventional setup focusing on isolated resonances in EP-related studies~\cite{Alu-19,ozdemir-19,EI-18}, the collective states arising from the collective interactions among these emitters provide a new platform for exploring non-Hermitian physics. These interference-enabled collective states, such as long-lived dark states formed via destructive interference, enable rich applications ranging from quantum memories to strong couplings in open environments~\cite{Dicke-54,Ziqi-22,Kannan-20,corzo-19,Tiranov-13,Zanner-22,Painter-19}. By engineering the radiative properties of individual emitters, these collective states can be manipulated to achieve a broad range of unconventional wave phenomena.

\begin{figure*}	
	\includegraphics[width=0.98\textwidth]{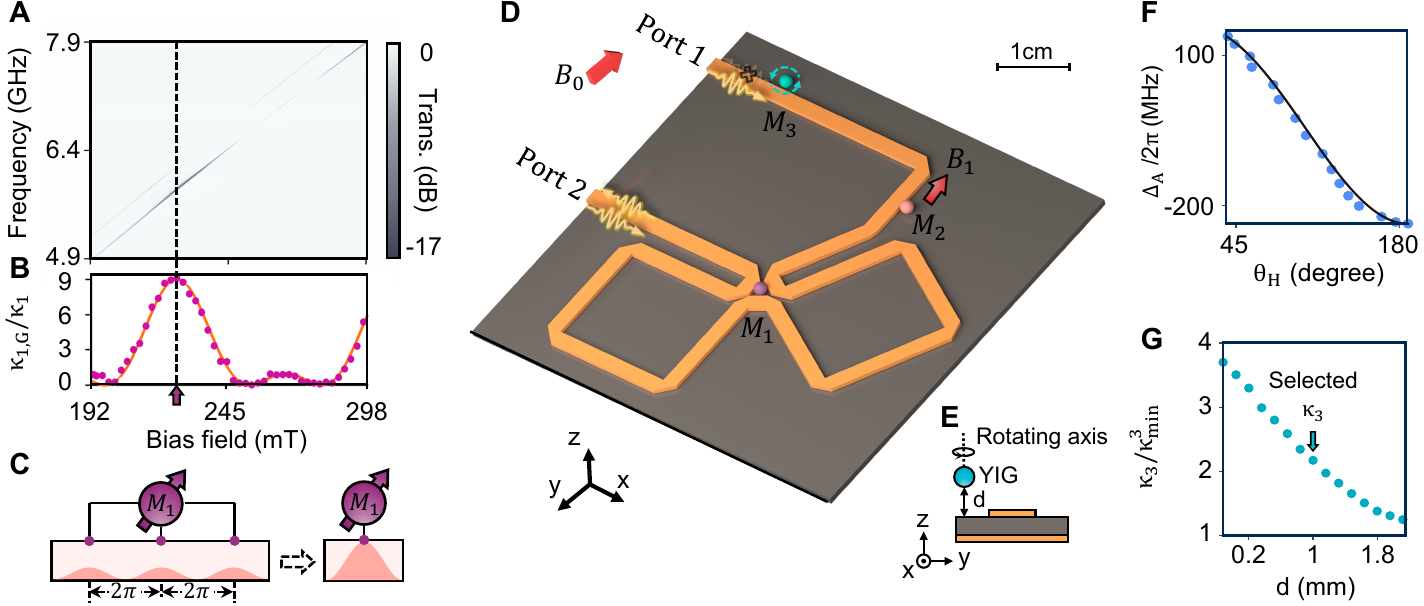}
	\caption {\textbf{Design of the asymmetric magnon mode array with giant spin ensemble.} 
		(\textbf{A}) Transmission mapping of $M_1$ as a function of the bias magnetic field, showing periodic modulation in dip depth due to phase-dependent coupling.
		(\textbf{B}) Extracted radiative damping rates of the giant spin ensemble (GSE) (dots) as a function of frequency, fitted using Eq.~(\ref{GSE}) (solid curve). 
		(\textbf{C}) Schematic of a GSE with three spatially separated coupling points, enabling collective constructive interference that significantly enhances the radiative coupling between magnon mode $M_1$ and the waveguide. 
		(\textbf{D}) Experimental setup: YIG spheres $M_1$ (purple) and $M_2$ (pink) are mounted on the waveguide (orange). The frequency of $M_2$ is tuned via a local field $B_1$ generated by a coil beneath it. Sphere $M_3$ (cyan) is attached to a cantilever connected to a rotation stage. All spheres are magnetized by a global field $B_0$.	
		(\textbf{E}) Side view showing the adjustable position of $M_3$ along the $x$ and $z$ axes. 
		(\textbf{F}) Tunability of $M_3$’s resonance frequency via anisotropy field control, plotted versus rotation angle $\theta_{\rm H}$.
		(\textbf{G}) Extracted radiative damping rate $\kappa_3$ of $M_3$ versus its vertical distance $d$ from the waveguide.  
	}
	\label{fig1}
\end{figure*}

In this work, we demonstrate that the combination of an RL EP and deliberate inversion-symmetry breaking in a collective magnonic system leads to \textit{unidirectional} reflectionless behavior with \textit{enhanced bandwidth}. Leveraging the flexible tunability of the magnonic system~\cite{Dengke-17,Yang-21,Jinwei-23,Xufeng-24,Yao-2023,Jinwei-2024,Peng-19,Qian-24,Dong-23,Wolski-20,Xia-18,Rameshti-22,Huebl-13,Shen-21,Shen-22,YiLi-2019,Yuan-22,Tabuchi-14,Xufeng-14,Tabuchi-15,Wang-18,Wang-2019,Harder-18,Tao-20,Xu-23,Dany-20,Bai-15,Xufeng-15,ziqi-25}, we design an inhomogeneous anti-Bragg magnon array composed of three spatially separated yttrium iron garnet (YIG) spheres coupled to a meandering waveguide. The Kittel mode of magnons (uniform spin precession mode) in each YIG sphere is coupled to the traveling photon mode in the waveguide. The system forms an effective magnonic mirror array (MMA), where two boundary magnonic mirrors create a dark-state cavity that coherently couples to the central magnon mode. To break the system’s inversion symmetry, we implement a giant spin ensemble (GSE) scheme~\cite{Ziqi-22} in which one of the magnonic mirrors is strongly coupled to the waveguide via three spatially separated points, leading to constructively enhanced radiative decay. In contrast, the other two YIG spheres couple to the waveguide at a single point each and can be regarded as two small spin ensembles. This asymmetric spatial configuration readily breaks the central inversion symmetry of the system.

In experiment, we observe that only one reflection direction exhibits the hallmark features of an RL EP, including reflection suppression with a flattened quartic lineshape and significantly broadened bandwidth. The other direction remains highly reflective, confirming the unidirectional nature of the system. Moreover, by tuning the magnon frequencies, we continuously track the coalescence and splitting of reflectionless states, revealing their coherent origin and strong sensitivity to collective dynamics. Our results showcase a new route for engineering non-Hermitian phenomena in collective systems and open the door to broadband, unidirectional wave control in both classical and quantum regimes.

\begin{figure}
	\includegraphics[width=0.68\textwidth]{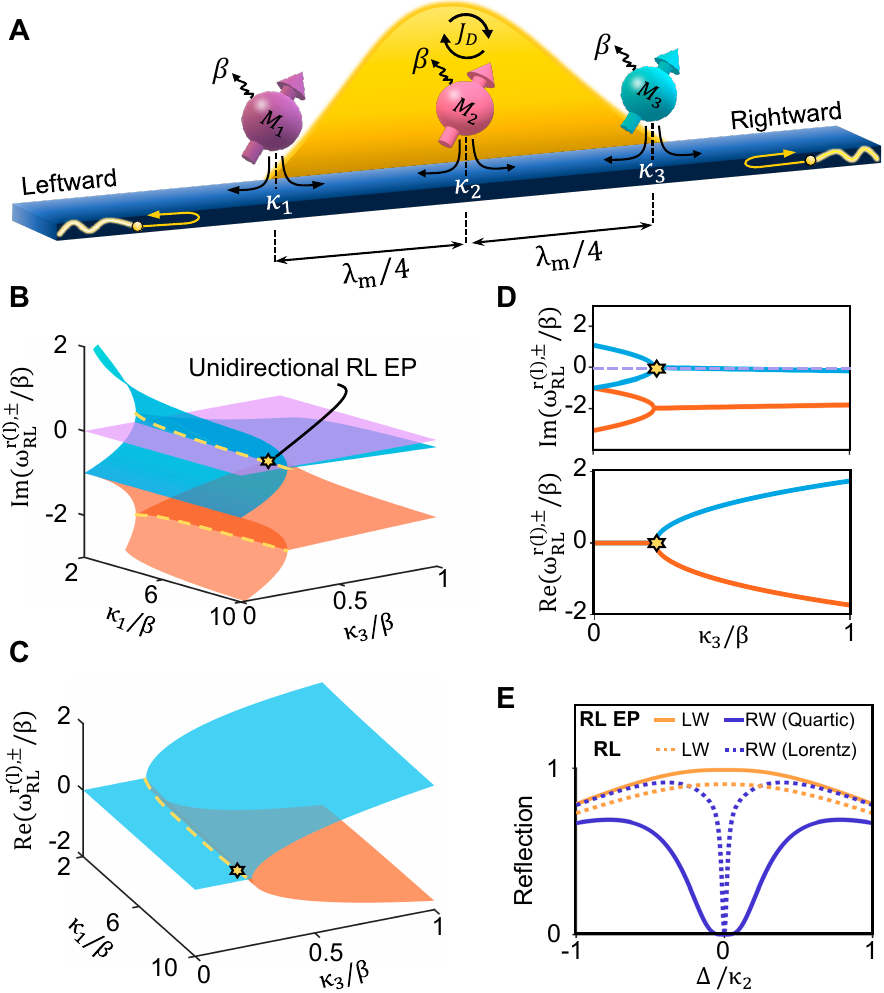}
	\centering
	\caption {\textbf{Unidirectional exceptional point of reflectionless states.} 
		(\textbf{A}) Schematic of the magnon mode array (MMA), where three magnon modes are coupled to a waveguide with external (intrinsic) dissipation rates $\kappa_{i}$ ($\beta$). Adjacent magnon modes are spaced by $\lambda_{\rm{m}}/4$. 
		(\textbf{B} and \textbf{C}) Imaginary (B) and real (C) parts of the RL eigenfrequencies for rightward and leftward incidence, plotted as functions of the magnon radiative decay rates $\kappa_{\rm{1}}$ and $\kappa_{\rm{3}}$. Yellow dashed lines indicate the EP conditions. Only the imaginary parts of the RL eigenfrequencies for rightward incidence intersects the zero-imaginary-frequency (purple) plane, marking the unidirectional RL EP. Here, $\kappa_2/\beta = 0.93$. 
		(\textbf{D}) Real and imaginary parts of the RL eigenfrequencies versus $\kappa_{\rm{3}}$, extracted from side views of panels B and C.
		(\textbf{E}) Reflection spectra for both incidence directions under conventional RL (dashed) and RL EP (solid) conditions. Orange and blue curves correspond to leftward and rightward incidence, respectively.
	}
	\label{fig2}
\end{figure}	

\subsection*{Results}
\subsubsection*{System and concepts}
Our MMA device is schematically illustrated in Fig.~\ref{fig1}D. It consists of a meandering waveguide and three 1 mm-diameter single-crystal YIG spheres. The considered Kittel mode in each sphere is denoted as $M_{\rm i}$, $i=1,2,3$, which has a total decay rate $\gamma_{\rm{i}}=\kappa_{i}+\beta$, where $\kappa_{i}$ and $\beta$ are the radiative and intrinsic losses, respectively. We apply a uniform bias magnetic field along the $y$-axis to ensure homogeneous magnetization of each YIG sphere. Once magnetization is saturated, the magnon resonance frequency is given by $\omega_{\rm{m}} = \gamma(H_{\rm{e}} + H_{\rm{A}})$, where $\gamma/2\pi = 28$~GHz/T is the gyromagnetic ratio, $H_{\rm{e}}$ is the bias field, and $H_{\rm{A}}$ denotes the anisotropy field. Notably, magnon mode 1 ($M_{1}$, purple) is coupled to the waveguide at three well-separated mitered corners that are equidistant, resulting in identical local interaction strengths $g_1$ between the magnon mode and the waveguide at each point. Along the waveguide, the separation between these coupling points exceeds the microwave wavelength, so the dipole approximation no longer applies. Consequently, interference among the coupling points becomes significant and regulates the properties of the GSE. By designing the coupling point separation $L$ to be equal, the effective radiative damping rate of the GSE, $\kappa_{\rm{1,G}}$, can be expressed as (see Supplementary Materials for details)
\begin{equation} 
	\label{GSE}
	\kappa_{\rm{1,G}}=\kappa_{\rm{1}}(3+4\cos\varphi_{\rm{G}}+2\cos2\varphi_{\rm{G}}),
\end{equation}
where $\varphi_{\rm{G}}=\omega_{\rm{1}}L/v$ is the accumulated phase between adjacent coupling points. This effective radiative damping rate $\kappa_{\rm{1,G}}$ can be continuously tuned by varying the magnon resonance frequency, as demonstrated by the dip along the diagonal of the measured transmission spectra versus bias magnetic field shown in Fig.~\ref{fig1}A. The depth and width of the transmission dip along the diagonal changes periodically with the resonance frequency, indicating modulation of the magnon-waveguide coupling. These variations are quantitatively extracted by fitting $\kappa_{\rm{1,G}}$, as presented in Fig.~\ref{fig1}B. This observation confirms the successful realization of a GSE~\cite{Ziqi-22}. When the latter two cosine terms in Eq.~(\ref{GSE}) simultaneously reach their maxima, the magnon-waveguide interactions at the three coupling points of $M_1$ interfere constructively, leading to a ninefold enhancement in $\kappa_{\rm{1,G}}$ relative to the single-coupling-point value $\kappa_{\rm{1}}$. In this configuration, the GSE can be simplified as a single-coupling-point emitter with enhanced and tunable radiative decay, as illustrated in Fig.~\ref{fig1}C.

More importantly, the introduction of the GSE significantly enhances the asymmetry of the system, facilitating the subsequent observation of the unidirectional RL states. The realization of these RL states relies on the collective behaviors of the MMA. In this work, we focus on the anti-Bragg condition, where the adjacent magnon modes are spaced by a distance of $\lambda_{\rm{m}}/4$ (Fig.~\ref{fig2}A). Here $\lambda_{\rm{m}}/2\pi = v/\omega_{\rm{m}}$ is the wavelength of the propagating field at the magnon resonance frequency $\omega_{\rm{m}}$, and $v$ denotes the photon group velocity. Under low excitation conditions, the effective Hamiltonian describing this magnonic array is given by

\begin{equation}
	H_{\text {eff}}/\hbar=\left(\begin{array}{ccc}
		\omega_{\rm{m}}-i\gamma_{\rm{1}} & J_{\rm{12}} & i \Gamma_{\rm{13}} \\
		J_{\rm{12}} & \omega_{\rm{m}}-i\gamma_{\rm{2}} & J_{\rm{23}} \\
		i \Gamma_{\rm{13}} & J_{\rm{23}} & \omega_{\rm{m}}-i\gamma_{\rm{3}}
	\end{array}\right),
	\label{effective_Hamiltonian}
\end{equation}
where $J_{sp} = \sqrt{\kappa_{s} \kappa_{p}}$ and $\Gamma_{sp} = \sqrt{\kappa_{s} \kappa_{p}}$ ($s,p=1,2,3$, $s \neq p$) represent the waveguide-mediated coherent and dissipative coupling strengths, respectively, between magnon modes $s$ and $p$. To characterize the system’s collective response through spectroscopic reflection measurements, we derive expressions for the rightward and leftward reflection coefficients (see Supplementary Materials for details):

\begin{equation}
	\scalebox{1.2}{$R^{r(l)}=\frac{(\Delta+i\beta)^{2}(\kappa_{1}-\kappa_{2}+\kappa_{3})-4 \kappa_{1} \kappa_{2} \kappa_{3} \pm \delta }{\text{det}(\omega\hat{I}-H_{\text {eff}})},$}
	\label{refectioncoefficient}
\end{equation}
where $\delta = 2i\kappa_{2}(\Delta + i\beta)(\kappa_{1} - \kappa_{3})$, $\Delta = \omega - \omega_{\rm{m}}$ is the detuning between the probe field and magnon mode, $\omega$ is the probe frequency, and $\hat{I}$ is the identity matrix. The key to realizing asymmetric reflection lies in the discriminant term $\delta$, which is only nonzero when the radiative damping rates of the boundary magnon modes are unequal, indicating that the system no longer preserves inversion symmetry. This reflection asymmetry becomes pronounced when the difference between $\kappa_1$ and $\kappa_3$ is significant. The implementation of the GSE readily fulfills this requirement.

\begin{figure*}
	\includegraphics[width=0.99\textwidth]{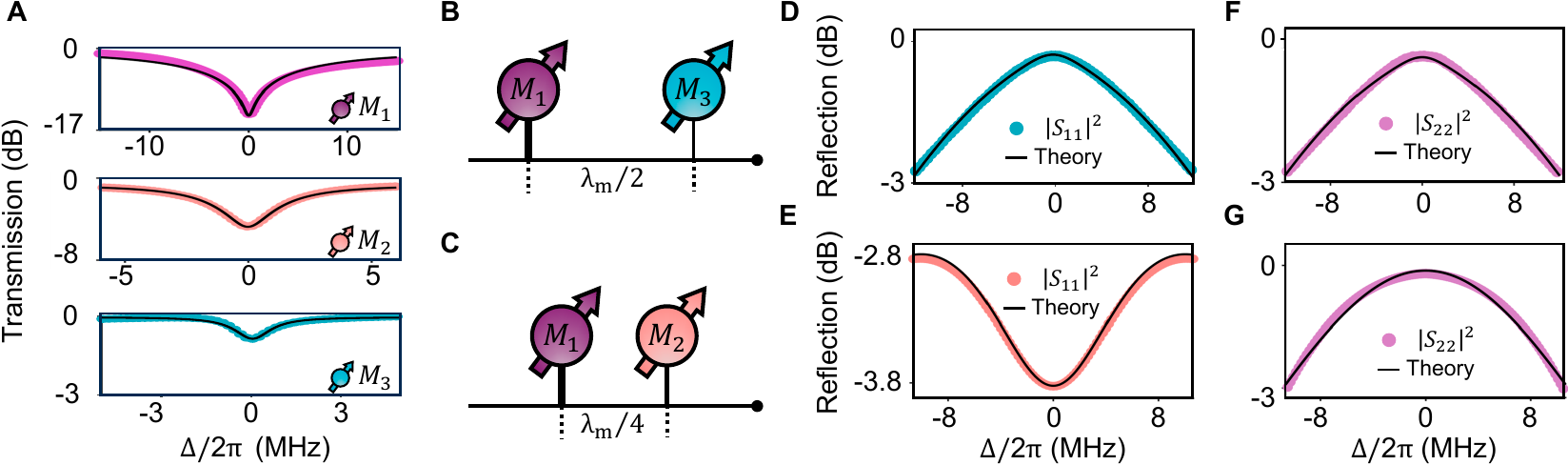}
	\caption {\textbf{Characterization of relative phases between magnon modes via reflection measurement.} 
		(\textbf{A}) Individually measured transmission spectra of the three magnon modes (dots) at the GSE resonance frequency, with theoretical fits (black curves).
		(\textbf{B} and \textbf{C}) Schematic illustrations of the relative phase configurations: (B) between $M_{\rm{1}}$ and $M_{\rm{3}}$, showing dissipative coupling with a $\pi$ phase difference; (C) between $M_{\rm{1}}$ and $M_{\rm{2}}$, showing coherent coupling with a $\pi/2$ phase difference.  
		(\textbf{D} and \textbf{F}) Reflection spectra measured from port 1 (D) and port 2 (F) when only $M_{\rm{1}}$ and $M_{\rm{3}}$ are present and dissipatively coupled.  
		(\textbf{E} and \textbf{G}) Reflection spectra measured from port 1 (E) and port 2 (G) when only $M_{\rm{1}}$ and $M_{\rm{2}}$ are present and coherently coupled.  
		All measurements were performed with only the corresponding pair of magnon modes present in the system.
	}
	\label{fig3}
\end{figure*}	

To realize the RL state, the numerator of Eq.~(\ref{refectioncoefficient}) must vanish, yielding two complex-valued solutions, $\omega_{\rm{RL}}^{r,\pm}$ and $\omega_{\rm{RL}}^{l,\pm}$, corresponding to the reflectionless conditions for rightward and leftward incidence, respectively. Fixing the radiative decay rate $\kappa_{2}$ of the central magnon mode $M_2$, we map the imaginary and real components of $\omega_{\rm{RL}}^{r(l),\pm}$ as functions of $\kappa_{1}$ and $\kappa_{3}$, presented as 3D surfaces in Figs.~\ref{fig2}B and C. The degeneracy of $\mathrm{Re}(\omega_{\rm{RL}}^{r,\pm})$ and $\mathrm{Re}(\omega_{\rm{RL}}^{l,\pm})$  in Fig.~\ref{fig2}C results in two distinguishable Riemann sheets. Along the $\kappa_{3}$ axis, these sheets can coalesce further. Despite this spectral degeneracy, the imaginary parts $\mathrm{Im}(\omega_{\rm{RL}}^{r(l),\pm})$ differ significantly due to the asymmetry in magnon damping rates across the system. The yellow dashed lines on the surfaces in Fig.~\ref{fig2}B and C indicate the EP line of both directions. For these RL states to be experimentally observable, the corresponding solution should be real, i.e., $\mathrm{Im}(\omega_{\rm{RL}}^{r(l),\pm}) = 0$. To identify such conditions, we introduce a zero plane (purple surface) in Fig.~\ref{fig2}B to locate intersections with the Riemann surfaces. Strikingly, only when $\kappa_1$ is sufficiently large (e.g., $\kappa_{1}/\beta \approx 10$) can the rightward solutions alone intersect this plane, indicating the emergence of unidirectional RL EP (yellow star) in our system. The side views in Fig.~\ref{fig2}D further confirm the presence of unidirectional RL EP. It should be noted that the frequencies $\omega_{\rm{RL}}^{r(l)}$ also correspond to the eigenvalues of the RL Hamiltonian $H_{\rm{RL}}^{r(l)}$~\cite{AnZ-24,Ferise-22,Han-23,Yang-21,Rotter-24}. The emergence of a unidirectional RL EP in our system essentially stems from $H_{RL}^{r}$ being PT-symmetric, while $H_{RL}^{l}$ is not (see Supplementary Materials for details). We compare the corresponding reflection spectra at this RL EP with those of a conventional RL state. As shown in Fig.~\ref{fig2}E, the quartic flattening of the reflection profile at the RL EP leads to a broadened unidirectional bandwidth. Note that the parameters used in Figs. \ref{fig2}B and \ref{fig2}C can be experimentally obtained by individually fitting the transmission response of each magnon mode. Moreover, they can be actively tuned in our fabricated device, as discussed below.

\subsubsection*{Realization of the asymmetric magnonic mirror array via a GSE}
A strongly enhanced radiative coupling in one of the magnonic mirror modes serves as a necessary precondition for realizing the unidirectional RL EP. However, this remains a major challenge to achieve in conventional magnonic systems~\cite{Ziqi-22,Wang-20,Yao-2023,Jinwei-23,Xufeng-24,Jinwei-2024,YiLi-2019,Yang-21}. This limitation underscores the crucial role of the GSE introduced in our work, which offers a unique method for fulfilling this condition. At the constructive interference frequency ($\omega_{\rm{1}}/2\pi = 5.9$~GHz), the significantly enhanced radiative coupling produces a broad Lorentzian transmission spectrum, as shown in the top panel of Fig.~\ref{fig3}A. The fitted radiative damping rate of $M_1$ is $\kappa_{\rm{1,G}}/2\pi = 13.23$~MHz. The individually measured transmission spectra for the other configured two magnon modes, $M_2$ (pink) and $M_3$ (green), are also shown in Fig.~\ref{fig3}A, with fitted radiative damping rates $\kappa_{\rm{2}}/2\pi = 0.76$~MHz and $\kappa_{\rm{3}}/2\pi = 0.19$~MHz, respectively. The intrinsic damping rates of all the three modes are fitted to $\beta/2\pi = 1.02$~MHz. The reduction of $\kappa_{\rm{3}}$ is clearly identified by a shallower resonance dip, which is controlled by displacing the corresponding sphere away from the microstrips (Fig.~\ref{fig1}G, see Methods for details). This MMA configuration meets the requirement for RL EP formation. Next, we adjust the relative positions of these spheres so that they can meet the anti-Bragg condition.

To explore the collective behaviors of the three magnon modes, their frequencies should be tuned into mutual resonance. For $M_2$, this is achieved by applying a local magnetic field $H_2$ via a coil beneath the sample. For $M_3$, its resonance frequency can also be individually regulated by harnessing the magnetocrystalline anisotropy field (Fig.~\ref{fig1}F, see Methods for details). This versatile tunability enables the realization and comprehensive characterization of the MMA system and its RL EPs.

As the magnon modes are collectively coupled to the waveguide, traveling photons mediate long-range interactions among them. To satisfy the anti-Bragg condition for the magnon array at the GSE resonance frequency $\omega_{1}$, the accumulated propagation phase $\varphi_{mn}$ between the magnon modes $M_{\rm{m}}$ and $M_{\rm{n}}$ should be precisely controlled by adjusting the spacing between the corresponding YIG spheres via step motors. In our implementation, magnon modes $M_2$ and $M_3$ are individually coupled to the GSE magnon mode $M_1$ at 5.9~GHz. To ensure the stable emergence of the RL EP, the magnon resonance frequency is chosen to avoid perturbations from higher‑order magnetostatic modes, and the input power is -20 dBm to suppress Kerr‑nonlinearity‑induced frequency shifts.

As illustrated in Fig.~\ref{fig3}B, the magnon modes $M_1$ and $M_3$, serving as the magnonic mirrors of the MMA, are separated by a $\pi$ phase shift, forming a dissipatively coupled configuration. Despite the strong inversion asymmetry between these two mirrors, the reflection spectra in both directions, shown in Figs.~\ref{fig3}D and F, appear nearly identical. This arises because only the bright eigenmode is accessible in the reflection measurements, producing a broad, high-reflectivity peak in both directions. In contrast, when the probing magnon mode $M_2$ is positioned with a $\pi/2$ phase shift relative to $M_1$, coherent coupling occurs, as shown in Fig.~\ref{fig3}C. In this case, the system asymmetry becomes clearly visible in the reflection spectra (Figs.~\ref{fig3}E and G), as both eigenstates contribute to the observed signal. By adjusting the relative positions of the three YIG spheres, we realize an inhomogeneous magnon mode array that can be interpreted as a probing magnon mode $M_2$ coupled to a cavity mode supported by the magnonic mirrors, as schematically illustrated in Fig.~\ref{fig4}A. The cavity mode corresponds to the dark mode formed by the dissipatively coupled magnon modes $M_1$ and $M_3$, and is decoupled from the waveguide except through its intrinsic damping rate $\beta$. The effective coupling strength between $M_2$ and the magnonic cavity mode (dark mode) is
\begin{equation}
	J_{\rm{D}} = \frac{2\sqrt{\kappa_{\rm{1}}\kappa_{\rm{2}}\kappa_{\rm{3}}}}{\sqrt{\kappa_{\rm{1}}+\kappa_{\rm{3}}}}.
\end{equation}
Although the asymmetry between $M_1$ and $M_3$ induces an additional coupling channel between $M_2$ and the bright mode, the bright mode has a significantly larger linewidth due to its stronger radiative damping. As a result, its quality factor is always lower than that of the dark mode, so its influence can be reasonably ignored in spectroscopic measurements.

\subsubsection*{Unidirectional EP of reflectionless states}
\begin{figure}
	\includegraphics[width=0.99\textwidth]{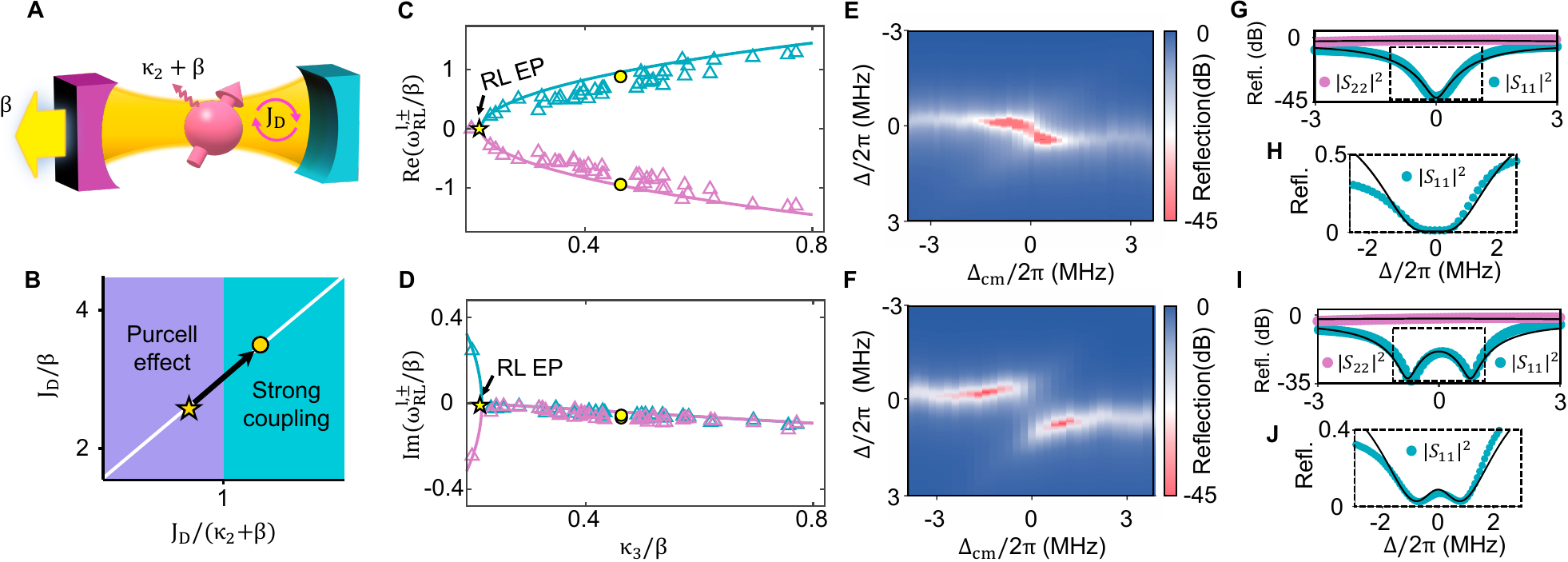}
	\centering
	\caption {\textbf{Unidirectional detection in the effective magnonic mirror cavity system.}  
		(\textbf{A}) Schematic illustration of the effective magnonic mirror cavity system, where a probing magnon mode couples to a dark cavity mode formed by two dissipatively coupled magnonic mirrors.  
		(\textbf{B}) Parameter space showing two coupling regimes, separated by the ratio between the coupling strength $J_{\rm D}$ and the linewidths of the cavity and magnon modes. The white line tracks the system's trajectory as $\kappa_3$ is tuned. Stars and circles mark the cases in panels E and F.
		(\textbf{C} and \textbf{D}) Fitted real (C) and imaginary (D) parts of the RL-state eigenfrequencies $\omega_{\rm{RL}}^{l,\pm}$ versus $ \kappa_{3}/\beta$.
		(\textbf{E} and \textbf{F}) Reflection mappings of $|S_{11}|^2$ measured by gradually rotating sphere $M_3$ in two different coupling regimes: $\kappa_3/2\pi = 0.19$~MHz (E), and $0.55$~MHz (F).  
	(\textbf{G} and \textbf{I}) Reflection spectra measured at resonance. Cyan and purple dots represent $|S_{11}|^2$ and $|S_{22}|^2$ extracted from E and F, respectively. Black curves show theoretical fits.
		(\textbf{H} and \textbf{J}) The $|S_{11}|^2$ reflection spectra in (G) and (I) plotted on a linear scale.
	}
	\label{fig4}
\end{figure}

Following the framework of cavity magnonics~\cite{Xufeng-14}, we can classify the coupling regimes of the MMA system based on the ratio of coupling strength to dissipation, as plotted in Fig.~\ref{fig4}B. Since $J_{\rm{D}}$ depends on the radiative decay rates of $M_1$ and $M_3$, we can continuously tune the system along the white trajectory in Fig.~\ref{fig4}B by adjusting $\kappa_{3}$. With this tunability, we next demonstrate the experimental realization of the unidirectional RL EP.

To clearly observe the unidirectional RL EP and the coupling behaviors between the magnonic mirror cavity and the probing magnon mode, we fix the global magnetic field and bring the $M_1$ and $M_2$ modes into resonance via a local field $H_1$. We then gradually rotate the YIG sphere 3 to effectively vary the frequency detuning $\Delta_{\rm{cm}}$ between the magnonic mirror cavity mode and the probing magnon mode. The RL EP condition is verified by extracting the real and imaginary parts of the RL-state eigenvalues $\omega_{\rm{RL}}^{l,\pm}$, obtained by directly fitting the reflection spectra $|S_{11}|^2$ when the magnonic mirror cavity mode and the probing magnon mode are tuned into resonance. When the real part $\mathrm{Re}(\omega_{\rm{RL}}^{l,\pm})$ and imaginary part $\mathrm{Im}(\omega_{\rm{RL}}^{l,\pm})$ coalesce simultaneously, the system reaches an EP. By tuning $\kappa_3$, we fit a series of reflection spectra to obtain the continuous variation in the eigenvalues, as shown in Figs.~\ref{fig4}C and \ref{fig4}D, where the emergence of the RL EP is clearly visible at $\kappa_{\rm{3}}/2\pi = 0.19$~MHz. The RL EP condition is located in the Purcell coupling regime, as marked by a star point in Fig.~\ref{fig4}B. It should be noted that $|S_{22}|^2$ remains a broad reflection peak throughout the variation of $\kappa_{3}$. As a result, fitting $|S_{22}|^2$ spectrum to extract the real and imaginary parts of $\omega_{\rm{RL}}^{r,\pm}$ is not feasible, since the resonance features are not sufficiently resolved in the measured spectra~\cite{Han-23} (see Supplementary Material for details). These results agree well with the theoretical predictions demonstrated in Fig.~\ref{fig2}D. At the RL EP, the coupling strength matches the linewidth of $M_2$, preventing spectral splitting of the hybridized modes. This is identified in the reflection spectra measured through waveguide port one (Fig.~\ref{fig4}E), where a broadened spectrum is observed at resonance.

Figure~\ref{fig4}G presents the reflection spectra measured at resonance for signals incident from both ports, revealing pronounced directional behavior. A clear unidirectional response is observed, with the reflection from port one exhibiting an almost reflectionless dip ($|S_{11}|^2$), while the reflection from port two remains near unity. It is worth noting that this result shows a flattened quartic lineshape in the linear-scale plot in  Fig.~\ref{fig4}H, which is a distinct signature of the EP. The mechanisms can be understood as follows. When excitation is launched from the high-reflectivity magnonic mirror ($M_1$, i.e., port two), the signal is predominantly reflected and fails to access the MMA. Conversely, when incident from the low-reflectivity magnonic mirror ($M_3$, i.e., port one), the microwave signal couples into the MMA and interacts with the probing magnon mode ($M_2$), producing a distinct RL valley. 

The RL EP behavior is further validated by enhancing the coupling strength to shift the system into the strong coupling regime. As shown in Fig.~\ref{fig4}F, when the magnonic mirror cavity mode resonates with the probing magnon mode, the hybridized polariton modes exhibit a clear spectral splitting. The system's unidirectionality is preserved and two well-separated Lorentzian dips are observed in the port-one reflection spectrum at resonance (Fig.~\ref{fig4}I). The bandwidth of one Lorentizian dip is substantially narrower than that of the quartic lineshape at the RL EP (Fig.~\ref{fig4}J), leading to the reduction of the unidirectional reflectionless bandwidth. The asymmetry in the reflection spectra on a linear scale in Figs.~\ref{fig4}H and \ref{fig4}J arises from the Fano interference introduced by the GSE. Importantly, this effect does not compromise the appearance of the quartic dip. While such mode splitting has been observed previously~\cite{Painter-19}, the non-Hermitian asymmetry engineered in our system enables its observation from a single direction. This unique unidirectional accessibility opens up new possibilities for developing direction-selective photonic devices in open waveguide systems. 

\subsection*{Discussion}
In conclusion, we have proposed and experimentally demonstrated a unidirectional reflectionless exceptional point (RL EP) in a magnonic mirror array (MMA) system, enabled by a collective dark state. A giant spin ensemble (GSE) with three spatially separated coupling points is employed as a tunable magnonic mirror, providing an effective method to control the system’s asymmetry. This configuration results in a substantial enhancement of the magnon’s radiative damping and enables the emergence of unidirectional reflectionless behavior. By finely tuning the detuning and coupling strengths of the magnon modes, we observe the coalescence of RL eigenfrequencies in one direction, manifested as a flattened quartic reflection dip, which is an unambiguous signature of the RL EP. This distinctive lineshape is beneficial for applications such as stealth and energy harvesting technologies~\cite{Rotter-24}.

Interestingly, the evolution of unidirectional RL states also indicates coherent interactions among the hybridized modes in the MMA system.  The reflectionless dip in the $|S_{11}|^2$ reflection spectrum bifurcates beyond the EP as the coupling strength $J_{D}$ increases, signifying the transition of the equivalent magnonic mirror cavity system into the strong-coupling regime. The unidirectional RL state can thus provide a unique opportunity to directly observe MMA polariton behaviors via waveguide transmission in a single direction. This directionality persists until the system regains inversion symmetry (see Supplementary Materials for details).  

This work not only enriches the understanding of RL EPs in non-Hermitian asymmetric systems, but also establishes a GSE-based reconfigurable and scalable platform for manipulating directional wave control, thereby broadening the scope of GSE applications. The GSE can be readily implemented on lithographic platforms using broadband, high transmission microstrip structures. By adopting analogous meandering waveguide strategies~\cite{Song-22} and appropriate emitters (e.g., excitation modes in antiferromagnets~\cite{Yamaguchi-10,Seifert-18} and quantum dots~\cite{Lodahl-15,Arcari-14}), the approach can naturally be extended to the terahertz and optical frequency domains. The demonstrated architecture is also compatible with integrated microwave photonics and can be scaled to complex magnonic lattices or hybrid quantum systems.

\subsection*{Materials and Methods}
\textbf{Device design}\\
The meandering waveguide shown in Fig.~\ref{fig1}D is fabricated on a 0.813 mm thick RO4003C substrate with a 1-oz copper conductor. In order to achieve the GSE with three coupling points, the microstrip with a width of 1.82 mm is connected with miter corners. The 1 mm diameter YIG spheres that we use are commercially available.\\
\textbf{Measurement and Parameter Tuning}\\The reflection are measured using a vector network analyzer (KEYSIGHT PNA-L Network Analyzer N5232B) with an input power of -10 dBm. The resonance frequency of the $M_3$ mode can be independently tuned by changing the anisotropic field, which depends on the angle $\theta_{\rm H}$ between the crystal axis and the applied field~\cite{Ziqi-22}. We implement this by mounting $M_3$ on a cantilever driven by a rotary motor, allowing angular adjustment via spatial translation. As shown in Fig.~\ref{fig1}F, this enables continuous tuning of $\omega_{3}/2\pi$ approximately over 300 MHz range. The regulation of $\kappa_{\rm{3}}$ is achieved by translating the green sphere along the $z$ axis (Fig.~\ref{fig1}E), with extracted results displayed in Fig.~\ref{fig1}G. Since $M_{1,2,3}$ are identical YIG spheres, their radiative damping rates into the waveguide should be equal under identical coupling conditions. In our setup, $M_{3}$ is intentionally placed farther from the microstrip, so its radiative damping rate is smaller than $M_{2}$. The $M_{1}$ is designed as a GSE to possess the greater radiation due to the constructive interference between the coupling points. Their difference is clearly evident in the corresponding transmission spectra.

\newpage


\clearpage 
\bibliography{refs}
\bibliographystyle{sciencemag}


\section*{Acknowledgments}
\paragraph*{Funding:}
J. Q. You acknowledges the financial support from the National Key Research and Development Program of China under grant no.~2022YFA1405200, the National Natural Science Foundation of China under grant no.~U25A20199 and no.~$92265202$, and the Science and Technology Project of Zhejiang Province under grant no.~$2025\rm{C}01028$. Y. P. Wang acknowledges the financial support from the National Key Research and Development Program of China under grant no.~2023YFA1406703, the National Natural Science Foundation of China under grant no.~$12174329$, the Fundamental Research Funds for the Central Universities under grant no.~$2024\rm{FZZX}02\rm{-}01\rm{-}02$ and the Zhejiang Provincial Natural Science Foundation of China  no.~$\rm{LR}26\rm{A}040001$. Z. Q. Wang acknowledges the financial support from the National Natural Science Foundation of China under grant no.~$123\rm{B}2064$.
\paragraph*{Author contributions:}
Z.Q.W., Y.P.P., Y.P.W. and J.Q.Y. initiated the research project, Z.Q.W., Y.P.P. designed the sample structure with input from Y.P.W. and J.Q.Y., Z.Q.W. and Y.P.P. realized the experiments and carried out the data analysis, Y.P.P., Z.Q.W., Y.P.W. and J.Q.Y. developed the theory, Z.Q.W., Y.P.P., Y.P.W., and J.Q.Y. drafted the manuscript, J.Q.Y. supervised the project.	
\paragraph*{Competing interests:}
There are no competing interests to declare.
\paragraph*{Data and materials availability:}All data needed to evaluate and reproduce the results in the paper are present in the paper and/or the Supplementary Materials.


\subsection*{Supplementary materials}
Supplementary Text\\
Figures S1 to S8\\


\newpage


\renewcommand{\thefigure}{S\arabic{figure}}
\renewcommand{\thetable}{S\arabic{table}}
\renewcommand{\theequation}{S\arabic{equation}}
\renewcommand{\thepage}{S\arabic{page}}
\setcounter{figure}{0}
\setcounter{table}{0}
\setcounter{equation}{0}
\setcounter{page}{1} 


\begin{center}
\section*{Supplementary Materials for\\ \scititle}


Zi-Qi Wang$^{1\ast}$,
Yuan-Peng Peng$^{1\ast}$,
Yi-Pu Wang$^{1\dagger}$,
J. Q. You$^{1,2\ddagger}$\\

\small$^\ast$ These authors contributed equally to this work. \\
\small$^\dagger$ Corresponding author. Email: yipuwang@zju.edu.cn \\
\small$^\ddagger$ Corresponding author. Email: jqyou@zju.edu.cn

\end{center}

\subsubsection*{This PDF file includes:}
Supplementary Text\\
Figures S1 to S8\\

\newpage

\subsection*{Supplementary Text}
\subsubsection*{S1. Exceptional Points of the Scattering Matrix and Reflectionless States}

A scattering exceptional point (EP) arises at specific locations in the parameter space where the non-Hermitian scattering matrix $S$ becomes defective, i.e., its eigenvalues and eigenvectors coalesce. For a reciprocal two-port system, the scattering matrix can be written as
\begin{equation}
	S = \begin{pmatrix}
		T_l & R_l \\
		R_r & T_r
	\end{pmatrix},
	\label{S}
\end{equation}
where $T$ and $R$ denote the transmission and reflection coefficients, respectively, with subscripts indicating the direction of signal incidence. For example, $R_l$ ($T_l$) represents the reflection (transmission) coefficient for a signal incident from the left port.

In a reciprocal system ($T_l = T_r = t$), the eigenvalues of the $S$-matrix are given by
\begin{equation}
	\omega_{S}^{\pm} = t \pm \sqrt{R_l R_r}.
\end{equation}
An EP is reached when the square-root term vanishes, i.e., when the product $R_l R_r = 0$. This condition typically yields multiple frequencies $\omega_{\mathrm{RL}}^{(1)}, \omega_{\mathrm{RL}}^{(2)}, \dots$ corresponding to reflectionless (RL) states. A further degeneracy of these RL states defines a reflectionless exceptional point (RL EP). Previous studies have primarily focused on RL EPs emerging in systems with inversion symmetry, where both reflection coefficients vanish simultaneously ($R_l = R_r = 0$), leading to bidirectional RL EPs. However, RL EPs can also occur in systems without inversion symmetry. For example, when $R_l = 0$ and $R_r \ne 0$, the system allows for the realization of \textit{unidirectional} RL EPs.

	\subsubsection*{S2. Effective Hamiltonian of a GSE-Waveguide System}
\label{Secton2}

In this section, we consider a more general case in which multiple giant spin ensembles (GSEs) interact with a meandering microwave microstrip waveguide through arbitrary coupling points, as illustrated in Supplementary Fig.~\ref{diagram}. The magnon mode we only concern in this work is the Kittel mode. The Kittel mode in each YIG sphere interacts with both right- and left-propagating photon modes in the waveguide. The total Hamiltonian of the GSE-waveguide system is given by $\hat{H} = \hat{H}_{\mathrm{wg}} + \hat{H}_{\rm{m}} + \hat{H}_{\mathrm{int}}$, where $\hat{H}_{\mathrm{wg}}$, $\hat{H}_{\rm{m}}$, and $\hat{H}_{\mathrm{int}}$ describe the waveguide, the magnon modes, and their mutual interaction, respectively. In the following analysis, we set $\hbar = 1$ for simplicity. We consider $N$ GSEs, where the $i$th GSE couples to the waveguide through $P_i$ spatially separated points. Under the rotating-wave approximation, the real-space Hamiltonian of the system reads~\cite{Roy-17,Fan-09,Fan-092}

\begin{figure*}[t]
	\centering
	\includegraphics[scale=0.55]{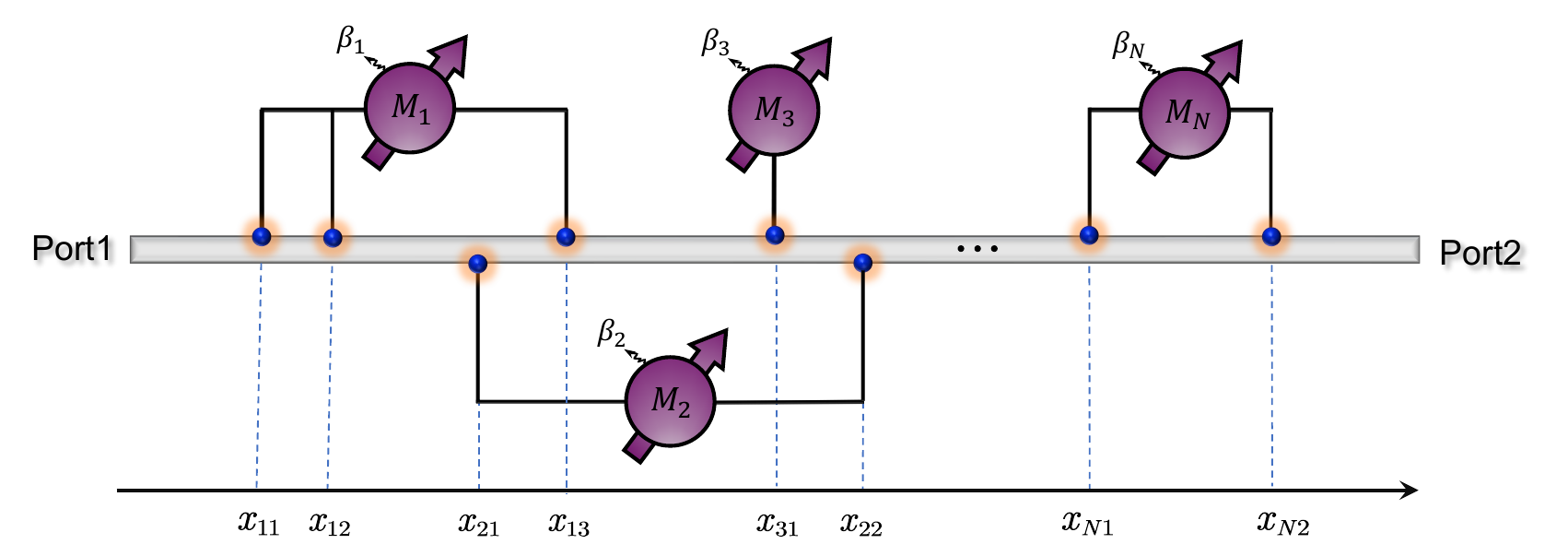}
	\caption{\textbf{Schematic of the GSE–waveguide system.} Multiple YIG spheres are coupled to the waveguide at arbitrary positions. The waveguide has two ports labeled Port 1 and Port 2.}
	\label{diagram}
\end{figure*}

\begin{equation}
	\hat{H}_{\rm{m}} = \sum_{i=1}^{N} (\omega_i - i \beta_i) \hat{m}_i^\dagger \hat{m}_i,
\end{equation}

\begin{equation}
	\hat{H}_{\mathrm{wg}} = -i v_{\mathrm{g}} \int dx \left[ \hat{c}_r^\dagger(x) \frac{\partial}{\partial x} \hat{c}_r(x) - \hat{c}_l^\dagger(x) \frac{\partial}{\partial x} \hat{c}_l(x) \right],
\end{equation}

\begin{equation}
	\hat{H}_{\mathrm{int}} = \int dx \sum_{i=1}^{N} \sum_{p=1}^{P_i} g_{ip} \delta(x - x_{ip}) \left[ \hat{c}_r^\dagger(x) \hat{m}_i + \hat{c}_l^\dagger(x) \hat{m}_i + \mathrm{H.c.} \right].
\end{equation}	
Here $\omega_i$ and $\beta_i$ denote the frequency and intrinsic loss rate of the Kittel mode in the $i$th YIG sphere, which follows $\omega_i = \gamma (H_{e,i} + H_A)$, where $\gamma/2\pi = 28~\mathrm{GHz/T}$ is the gyromagnetic ratio and $H_A$ is the anisotropy field. The operator $\hat{m}_i^\dagger$ ($\hat{m}_i$) creates (annihilates) a magnon excitation in the $i$th GSE. The operators $\hat{c}_r^\dagger(x)$ and $\hat{c}_l^\dagger(x)$ [$\hat{c}_r(x)$ and $\hat{c}_l(x)$] create (annihilate) right- and left-propagating photons at position $x$, respectively. The coordinate $x_{ip}$ denotes the position of the $p$th coupling point of the $i$th GSE and $v_{\mathrm{g}}$ is the photon group velocity in the waveguide. The parameter $g_{ip}$ denotes the coupling strength between the waveguide and the $i$th GSE at the $p$th coupling point.

In the single-excitation subspace, the eigenstate of the system can be written as
\begin{equation}
	|\Psi\rangle = \int dx \left[ \Phi_r(x) \hat{c}_r^\dagger(x) + \Phi_l(x) \hat{c}_l^\dagger(x) \right] |\emptyset\rangle + \sum_{i=1}^{N} f_i \hat{m}_i^\dagger |\emptyset\rangle,
	\label{IntEigenStat}
\end{equation}where $|\emptyset\rangle$ is the vacuum state (all modes unexcited), $\Phi_r(x)$ [$\Phi_l(x)$] is the wavefunction for a rightward (leftward) propagating photon and $f_i$ is the excitation amplitude of the magnon mode. Assume a single input photon with frequency $\omega = v_g k$ is incident from the left (Port 1), where $k$ is the momentum of the photon. The photon wavefunction ansatz reads

\begin{subequations}
	\begin{equation}
		\Phi_r(x) = e^{i k x} \left[ \theta(x_1 - x) + \sum_{s=1}^{N_c-1} t_s \theta(x - x_s)\theta(x_{s+1} - x) + t \theta(x - x_{N_c}) \right],
		\label{PhiR}
	\end{equation}
	\begin{equation}
		\Phi_l(x) = e^{-i k x} \left[ r \theta(x_1 - x) + \sum_{s=2}^{N_c} r_s \theta(x - x_{s-1})\theta(x_s - x) \right],
		\label{PhiL}
	\end{equation}
\end{subequations}where $x_s$ are the ordered positions of all coupling points, and $N_c = \sum_{i=1}^{N} P_i$. Here, $T_s$ ($R_s$) denotes the local transmission (reflection) amplitude in each segment, and $\theta(x)$ is the Heaviside function. By substituting Eq.~\eqref{IntEigenStat} into the Schr\"{o}dinger equation $\hat{H}|\Psi\rangle = \omega |\Psi\rangle$, and following the method in Ref.~\cite{peng-23}, we obtain the transmission and reflection coefficients

\begin{subequations}
	\begin{equation}
		T = 1 - i \mathbf{V}^\dagger (\omega \mathbf{I}- H_{\mathrm{eff}})^{-1} \mathbf{V},
		\label{tGeneral}
	\end{equation}
	\begin{equation}
		R = - i \mathbf{V}^{\top} (\omega\mathbf{I}- H_{\mathrm{eff}})^{-1} \mathbf{V}.
		\label{rGeneral}
	\end{equation}
\end{subequations}Here, $\mathbf{f} = (f_1, f_2, \dots, f_N)^{\top}$, and $\mathbf{V} = (\mathcal{V}_1, \mathcal{V}_2, \dots, \mathcal{V}_N)^{\top}$, where
\begin{equation}
	\mathcal{V}_i = \sum_{p=1}^{P_i} \sqrt{\kappa_{ip}} e^{i \varphi_{ip}},
	\label{vi}
\end{equation}
with $\kappa_{ip} = g_{ip}^2 / v_g$ being the decay rate into the waveguide, and $\varphi_{ip} = \omega_r x_{ip} / v_g$ is the phase at reference frequency $\omega_r$ (valid in the Markovian regime).

The effective non-Hermitian Hamiltonian is given by
\begin{equation}
	\mathcal{H}_{ij} = (\omega_i - i \beta_i) \delta_{ij} - i \sum_{p=1}^{P_i} \sum_{p'=1}^{P_j} \sqrt{\kappa_{ip} \kappa_{jp'}} e^{i |\varphi_{ip} - \varphi_{jp'}|}.
	\label{EffH}
\end{equation}Alternatively, it can be decomposed as
\begin{equation}
	\mathcal{H}_{ij} =
	\begin{cases}
		\omega_i + \Delta_{\mathrm{L},i} - i (\beta_i + \kappa_{\mathrm{eff},i}) & \text{for } i = j, \\
		J_{ij} - \frac{i}{2} \Gamma_{ij} & \text{for } i \ne j,
	\end{cases}
\end{equation}
where
\begin{subequations}
	\begin{equation}
		\Delta_{\mathrm{L},i} = \sum_{p,p'} \sqrt{\kappa_{ip} \kappa_{ip'}} \sin |\varphi_{ip} - \varphi_{ip'}|,
	\end{equation}
	\begin{equation}
		\kappa_{\mathrm{eff},i} = \sum_{p,p'} \sqrt{\kappa_{ip} \kappa_{ip'}} \cos(\varphi_{ip} - \varphi_{ip'}).
	\end{equation}
\end{subequations}
The coherent and dissipative couplings between distinct magnon modes are
\begin{subequations}
	\begin{equation}
		J_{ij} = \sum_{p}^{P_i} \sum_{p'}^{P_j} \sqrt{\kappa_{ip} \kappa_{jp'}} \sin |\varphi_{ip} - \varphi_{jp'}|,
	\end{equation}
	\begin{equation}
		\Gamma_{ij} = 2 \sum_{p}^{P_i} \sum_{p'}^{P_j} \sqrt{\kappa_{ip} \kappa_{jp'}} \cos(\varphi_{ip} - \varphi_{jp'}).
	\end{equation}
\end{subequations}
These results are general and apply not only to GSEs each with multiple coupling points but also to conventional spin ensembles where $P_i = 1$. All scattering spectra illustrated throughout this work are calculated using Eqs.~\eqref{tGeneral} and \eqref{rGeneral}.

\subsubsection*{S3. Effectively enhanced radiative decay rate of a GSE}

In this section, we consider the case in which only a single GSE or spin ensemble is coupled to the waveguide. This corresponds to $N=1$, and we denote the magnon frequency as $\omega_{\rm{m}}$. Equations~\eqref{vi} and \eqref{EffH} are simplified to

\begin{equation}
	V = \sum_{p=1}^{P} \sqrt{\kappa_p} e^{i \varphi_p},
\end{equation}

\begin{equation}
	H = \omega_{\rm{m}} + \Delta_{\mathrm{L}} - i (\beta + \kappa_{\mathrm{eff}}),
	\label{reEffH}
\end{equation}
where $P$ is the number of coupling points of the single GSE. If we assume identical radiative decay rates $\kappa_p = \kappa_0$ for all coupling points and a constant phase difference $\varphi_0$ between adjacent ones, i.e., $\varphi_p = (p-1)\varphi_0$, then we have
\begin{equation}
	\Delta_{\mathrm{L}} = \kappa_0 \sum_{p=1}^{P} \sum_{p'=1}^{P} \sin |(p - p') \varphi_0| 
	= \kappa_0 \frac{P \sin \varphi_0 - \sin(P \varphi_0)}{1 - \cos \varphi_0},
	\label{lamb-shift1}
\end{equation}

\begin{equation}
	\kappa_{\mathrm{eff}} = \kappa_0 \sum_{p=1}^{P} \sum_{p'=1}^{P} \cos[(p - p') \varphi_0] 
	= \kappa_0 \frac{1 - \cos(P \varphi_0)}{1 - \cos \varphi_0}.
	\label{kappa_eff1}
\end{equation}
According to Eq.~\eqref{kappa_eff1}, in the limit $\varphi_0 \rightarrow 2m\pi$ with $m \in \mathbb{Z}$, applying L'Hôpital's rule yields $\kappa_{\mathrm{eff}}=P^2\kappa_{\mathrm{0}}$. Thus, adding more coupling points provides an effective route to boost the spin ensemble's radiation.
By substituting these expressions into Eqs.~\eqref{tGeneral} and \eqref{rGeneral}, the transmission and reflection coefficients take the following forms:
\begin{subequations}
	\begin{equation}
		T = 1 - i \frac{\kappa_0 \sum_{p=1}^{P} \sum_{p'=1}^{P} e^{i (p' - p) \varphi_0}}{\omega - \omega_{\rm{m}} - \Delta_{\mathrm{L}} + i (\beta + \kappa_{\mathrm{eff}})} 
		= \frac{\omega - \omega_{\rm{m}} - \Delta_{\mathrm{L}} + i \beta}{\omega - \omega_m - \Delta_{\mathrm{L}} + i (\beta + \kappa_{\mathrm{eff}})},
		\label{t1}
	\end{equation}
	\begin{equation}
		R = -i \frac{\kappa_0 \sum_{p=1}^{P} \sum_{p'=1}^{P} e^{i (p' + p) \varphi_0}}{\omega - \omega_{\rm{m}} - \Delta_{\mathrm{L}} + i (\beta + \kappa_{\mathrm{eff}})} 
		= \frac{\kappa_{\mathrm{eff}} e^{-i 2\alpha}}{\omega - \omega_{\rm{m}} - \Delta_{\mathrm{L}} + i (\beta + \kappa_{\mathrm{eff}})},
		\label{r1}
	\end{equation}
\end{subequations}where the phase factor $\alpha$ satisfies
\begin{equation}
	\tan(2\alpha) = \frac{ \sum_{p, p'=1}^{P} \sin[(p + p')\varphi_0] }{ \sum_{p, p'=1}^{P} \cos[(p + p')\varphi_0] }.
\end{equation}
From Eqs.~\eqref{t1} and \eqref{r1}, it is clear that regardless of the number of coupling points, the transmission and reflection spectra of a single GSE always exhibit a Lorentzian lineshape centered at $\omega = \omega_{\rm{m}} + \Delta_{\mathrm{L}}$, with linewidth $\kappa_{\mathrm{eff}}$. Therefore, by tailoring the number of coupling points $P$ and the phase offset $\varphi_0$, the effective radiative decay rate $\kappa_{\mathrm{eff}}$ can be precisely engineered and significantly enhanced relative to $\kappa_0$, as illustrated in Supplementary Fig.~\ref{kappaeff}.

This result also indicates that a spatially distributed GSE with multiple coupling points can be reduced to a single spin ensemble. Since the main text focuses on spatially separated GSEs, in the following discussion, we treat each GSE as an spin ensemble with $P_i = 1$ for simplicity.

\begin{figure*}[t]
	\centering
	\includegraphics[scale=0.6]{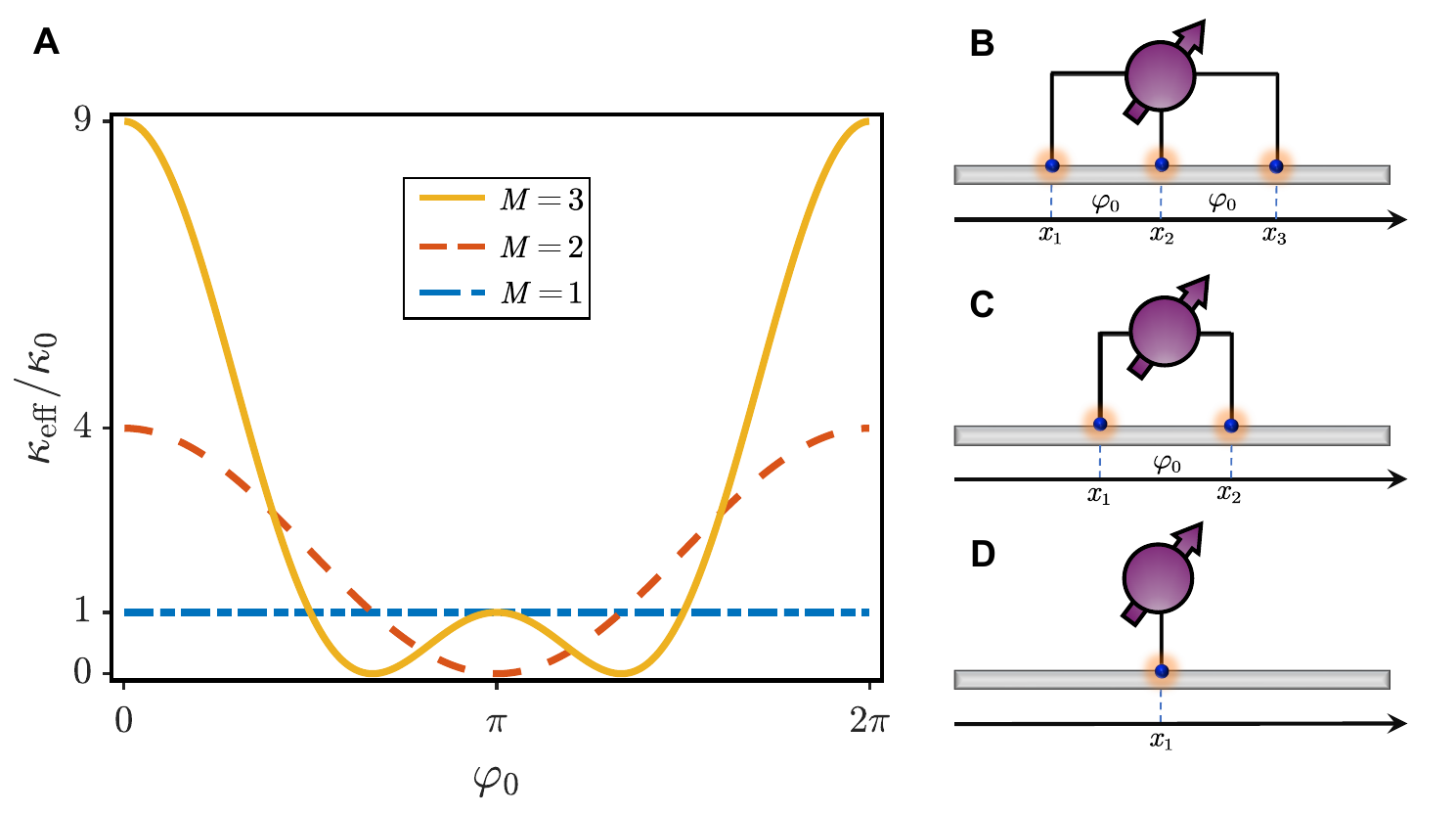}
	\caption{\textbf{Effective decay rate of a single GSE.} (\textbf{A}) Variation of the effective decay rate with phase difference between coupling points. Schematics for cases with (\textbf{B}) three, (\textbf{C}) two, and (\textbf{D}) one coupling points are shown.}
	\label{kappaeff}
\end{figure*}

\subsubsection*{S4. Reciprocal Transmission and Asymmetric Reflection}
\label{sec3}

From Eqs.~\eqref{tGeneral} and \eqref{rGeneral}, it is evident that the effective Hamiltonian $H_{\mathrm{eff}}$ governs the scattering behavior of the GSE-waveguide system. Being non-Hermitian, $H_{\mathrm{eff}}$ can be diagonalized as $H_{\mathrm{eff}} = \sum_n E_n |\Psi_n^R\rangle \langle \Psi_n^L|$, where $\langle \Psi_n^L | \Psi_{n'}^R \rangle = \delta_{nn'}$ defines the normalized biorthogonal basis. Consequently, the transmission and reflection amplitudes can be expressed in terms of the collective modes of the GSEs as

\begin{subequations}
	\begin{equation}
		T(\omega) = 1 - i \sum_{n=1}^{N} \frac{ \mathbf{V}^\dagger |\Psi_n^R\rangle \langle \Psi_n^L| \mathbf{V} }{ \omega - \text{Re}(E_n) - i \text{Im}(E_n) },
		\label{t_collective}
	\end{equation}
	\begin{equation}
		R(\omega) = -i \sum_{n=1}^{N} \frac{ \mathbf{V}^\top |\Psi_n^R\rangle \langle \Psi_n^L| \mathbf{V} }{ \omega - \text{Re}(E_n) - i \text{Im}(E_n) }.
		\label{r_collective}
	\end{equation}
\end{subequations}

These expressions show that the scattering process can be understood as an interference of multiple Lorentzian modes, each corresponding to a collective eigenmode of $H_{\mathrm{eff}}$ with center frequency $\text{Re}(E_n)$ and linewidth $\text{Im}(E_n)$. Each mode contributes a distinct scattering channel. The biorthogonal eigenstates of the non-Hermitian Hamiltonian are defined as
\begin{subequations}
	\begin{equation}
		H_{\mathrm{eff}} |\Psi_n^R\rangle = E_n |\Psi_n^R\rangle,
		\label{Rvector}
	\end{equation}
	\begin{equation}
		H_{\mathrm{eff}}^\dagger |\Psi_n^L\rangle = E_n^* |\Psi_n^L\rangle.
		\label{Lvector}
	\end{equation}
\end{subequations}
Due to the symmetry $H_{\mathrm{eff}} = H_{\mathrm{eff}}^\top$, it follows that $H_{\mathrm{eff}}^\dagger = (H_{\mathrm{eff}}^\top)^* = H_{\mathrm{eff}}^*$, and Eq.~\eqref{Lvector} becomes

\begin{equation}
	H_{\mathrm{eff}}^* |\Psi_n^L\rangle = E_n^* |\Psi_n^L\rangle.
\end{equation}By comparing with the complex conjugate of Eq.~\eqref{Rvector}, we obtain

\begin{equation}
	|\Psi_n^L\rangle = (|\Psi_n^R\rangle)^*, \quad \langle \Psi_n^L| = (\langle \Psi_n^R|)^*.
	\label{RtoL}
\end{equation}In Eqs.~\eqref{t_collective} and \eqref{r_collective}, the numerators are defined as coupling efficiency terms, which quantify the excitation efficiency of the collective eigenmodes of $H_{\mathrm{eff}}$ by the incident photon mode, i.e., the overlap between the input field and each eigenmode of the system~\cite{Nie-21}. Using Eq.~\eqref{RtoL}, the interaction spectrum can be expressed solely in terms of the right eigenstates

\begin{subequations}
	\begin{equation}
		\tilde{\eta}_n = \mathbf{V}^\dagger |\Psi_n^R\rangle \cdot \mathbf{V}^\top |\Psi_n^R\rangle,
		\label{teta}
	\end{equation}
	\begin{equation}
		\eta_n = \mathbf{V}^\top |\Psi_n^R\rangle \cdot \mathbf{V}^\top |\Psi_n^R\rangle.
		\label{eta}
	\end{equation}
\end{subequations}

To investigate the scattering properties, we consider photons incident from the right (Port 2), which corresponds to reversing the propagation direction ($k \rightarrow -k$), resulting in $\mathbf{V} \rightarrow \mathbf{V}^*$. The port-resolved scattering matrix elements are defined as

\begin{subequations}
	\begin{equation}
		T^l \equiv S_{21} = 1 - i \sum_{n=1}^{N} \frac{ \mathbf{V}^\dagger |\Psi_n^R\rangle \cdot \mathbf{V}^\top |\Psi_n^R\rangle }{ \omega - E_n } 
		\equiv 1 - i \sum_{n=1}^{N} \frac{ \tilde{\eta}_n^{(l)} }{ \omega - E_n },
		\label{s21}
	\end{equation}
	\begin{equation}
		T^r \equiv S_{12} = 1 - i \sum_{n=1}^{N} \frac{ \mathbf{V}^\top |\Psi_n^R\rangle \cdot \mathbf{V}^\dagger |\Psi_n^R\rangle }{ \omega - E_n } 
		\equiv 1 - i \sum_{n=1}^{N} \frac{ \tilde{\eta}_n^{(r)} }{ \omega - E_n },
		\label{s12}
	\end{equation}
	\begin{equation}
		R^l \equiv S_{11} = -i \sum_{n=1}^{N} \frac{ \mathbf{V}^\top |\Psi_n^R\rangle \cdot \mathbf{V}^\top |\Psi_n^R\rangle }{ \omega - E_n } 
		\equiv -i \sum_{n=1}^{N} \frac{ \eta_n^{(l)} }{ \omega - E_n },
		\label{s11}
	\end{equation}
	\begin{equation}
		R^r \equiv S_{22} = -i \sum_{n=1}^{N} \frac{ \mathbf{V}^\dagger |\Psi_n^R\rangle \cdot \mathbf{V}^\dagger |\Psi_n^R\rangle }{ \omega - E_n } 
		\equiv -i \sum_{n=1}^{N} \frac{ \eta_n^{(r)} }{ \omega - E_n }.
		\label{s22}
	\end{equation}
\end{subequations}
The superscripts indicate photons incident from the left and right, respectively. To avoid ambiguity, the transmission and reflection coefficients are redefined using uppercase letters. From these expressions, it is evident that $\tilde{\eta}_n^{(l)} \neq \tilde{\eta}_n^{(r)}$ or $\eta_n^{(l)} \neq \eta_n^{(r)}$ are necessary conditions for realizing nonreciprocal transmission or symmetric reflection. In the absence of time-reversal symmetry breaking, the coupling amplitudes from the left and right satisfy $\tilde{\eta}_n^{(l)} = \tilde{\eta}_n^{(r)}$, which leads to reciprocal transmission, i.e., $S_{21} = S_{12}$. However, for reflection, the interaction terms generally satisfy $\eta_n^{(l)} \neq \eta_n^{(r)}$, except in special cases where $\mathbf{V}^\dagger = \mathbf{V}^\top$ or the system exhibits spatial inversion symmetry. In these cases, symmetric reflection is recovered, while in all other situations, reflection asymmetry arises with $S_{11} \neq S_{22}$. In the following, we focus on analyzing the asymmetry in the reflection spectrum.

	\subsubsection*{S5. Reflectionless Hamiltonian of a two-port waveguide magnonic system}
	To reveal the condition for the coalescence of the RL states, the reflectionless Hamiltonian $H_{\rm RL}$ should be dervied analytically. We consider a $N\times N$ effective Hamiltonian $H_{\mathrm{eff}}$ with coupling vector $\mathbf{V}=(v_1,v_2,\dots,v_N)\in \mathbb{C}^{1\times N}$. The reflection coefficient is given by
	\begin{equation}
		R(\omega)=-i\,\mathbf{V}\,(\,\omega \mathbf{I}_N-H_{\mathrm{eff}}\,)^{-1}\mathbf{V}^{\top}. 
		\label{r_general2}
	\end{equation}
	Obviously, the above expression does not clearly indicate the condition under which the reflection coefficient vanishes. To resolve this issue, we transforms the reflection coefficient into a determinant ratio in which the zeros are dictated solely by the eigenvalues of the reflectionless Hamiltonian $H_{\rm RL}$. This formula is widely used in other studies on EP arising in scattering~\cite{Stone-20,Ferise-22,AnZ-24}, which provides a transparent way to uncover the physical mechanism and the conditions for degeneracy. 
	
	Because the expression in Eq.~\eqref{r_general2} differs from that of a one-port system~\cite{AnZ-24}, we cannot use a similar mathematical approach to obtain $H_{\rm RL}$ in a simple way. Here, we provide a method for solving the $H_{\rm RL}$ in a two-port system by performing an orthogonal transformation that aligns the vector $\mathbf{V}$ onto the last coordinate axis. Explicitly, there exists $\mathbf{S}\in O(N,\mathbb{C})$ such that
	\begin{equation}
		\mathbf{S}\,\mathbf{V}^{\top}=\sqrt{\mathbf{V}\mathbf{V}^{\top}}\,\mathbf{e}_N,
		\label{r_111}
	\end{equation}
	where $\mathbf{e}_N$ is the $N$th canonical basis vector. 
	An explicit construction is obtained by defining
	\begin{equation}
		\mathbf{w}=\mathbf{V}^{\top}-\sqrt{\mathbf{V}\mathbf{V}^{\top}}\,\mathbf{e}_N,
		\label{w}
	\end{equation}
	and taking
	\begin{equation}
		\mathbf{S}=\mathbf{I}_N-\frac{2\,\mathbf{w}\mathbf{w}^{\top}}{\mathbf{w}^{\top}\mathbf{w}}.
		\label{S}
	\end{equation}
	This $\mathbf{S}$ satisfies $\mathbf{S}^{\top}\mathbf{S}=\mathbf{I}_N$ and rotates $\mathbf{V}^{\top}$ exactly onto the last axis as required by Eq.~\eqref{r_111}.
	Writing $\mathbf{M}(\omega)=\omega \mathbf{I}_N-H_{\mathrm{eff}}$ and $\mathbf{M}'(\omega)=\mathbf{S}\,\mathbf{M}(\omega)\,\mathbf{S}^{\top}$, Eq.~\eqref{r_general2} becomes
	\begin{equation}
		R(\omega)=-i\,(\mathbf{V}\mathbf{V}^{\top})\,[\mathbf{M}'(\omega)^{-1}]_{NN}.
	\end{equation}
	From the adjugate formula,
	\begin{equation}
		[\mathbf{M}'(\omega)^{-1}]_{NN}=\frac{\det(\mathbf{M}'_{(N-1)\times(N-1)}(\omega))}{\det \mathbf{M}(\omega)}.
	\end{equation}
	Thus the reflection coefficient reduces to
	\begin{equation}
		R(\omega)=-i\,(\mathbf{V}\mathbf{V}^{\top})\,\frac{\det(\mathbf{M}'_{(N-1)\times(N-1)}(\omega))}{\det(\,\omega \mathbf{I}_N-\mathbf{H}_{\mathrm{eff}}\,)}.
	\end{equation}
	If $\mathbf{Q}\in\mathbb{C}^{N\times (N-1)}$ is an orthonormal basis with $\mathbf{V}\mathbf{Q}=0$, then
	\begin{equation}
		\mathbf{M}'_{(N-1)\times(N-1)}(\omega)=\mathbf{Q}^{\top}\mathbf{M}(\omega)\mathbf{Q}=\omega \mathbf{I}_{N-1}-\mathbf{Q}^{\top}\mathbf{H}_{\mathrm{eff}}\mathbf{Q}.
	\end{equation}
	Here $\mathbf{Q}$ can be chosen explicitly as the first $N-1$ columns of $\mathbf{S}^{\top}$, where $\mathbf{S}$ is the complex orthogonal matrix defined in Eq.~\eqref{r_111}. 
	This motivates the definition of the reflectionless Hamiltonian as
	\begin{equation}
		\mathbf{H}_{\mathrm{RL}}=\mathbf{Q}^{\top}\mathbf{H}_{\mathrm{eff}}\mathbf{Q}.
	\end{equation}
	Accordingly, the reflection coefficient can be expressed as
	\begin{equation}
		R(\omega)=-i\,(\mathbf{V}\mathbf{V}^{\top})\,\frac{\det(\omega \mathbf{I}_{N-1}-\mathbf{H}_{\mathrm{RL}})}{\det(\omega \mathbf{I}_N-\mathbf{H}_{\mathrm{eff}})}.
		\label{rl_ex}
	\end{equation}
	Hence the RL states are determined exactly by the eigenvalues of $\mathbf{H}_{\mathrm{RL}}$. It should be noted that this compact expression is valid only when $\mathbf{V}\mathbf{V}^{\top}\neq 0$. Therefore, $\mathbf{V}\mathbf{V}^{\top}=0$ does \emph{not} imply $R(\omega)\equiv 0$. If $\mathbf{V}\mathbf{V}^{\top}=0$, Eq.~\eqref{rl_ex} becomes ill-defined, but the original definition in Eq.~\eqref{r_general2} remains valid and generally yields nonzero values. 
	
	\subsubsection*{S6. Effective and Reflectionless Hamiltonian of an Anti-Bragg magnon array}
	To present the RL Hamiltonian of our system, we first derive the corresponding effective Hamiltonian. Base on the formula in Section S2, the effective Hamiltonian describing three magnon modes coupled to a waveguide can be written as
	\begin{equation}
		H_{\mathrm{eff}}=
		\begin{pmatrix}
			\omega_{\mathrm{m}} - i(\kappa_1 + \beta) &
			-i\sqrt{\kappa_1 \kappa_2}\, e^{i\frac{2\pi d_{12}}{\lambda_{\mathrm{m}}}} &
			-i\sqrt{\kappa_1 \kappa_3}\, e^{i\frac{2\pi d_{13}}{\lambda_{\mathrm{m}}}} \\[6pt]
			-i\sqrt{\kappa_1 \kappa_2}\, e^{i\frac{2\pi d_{12}}{\lambda_{\mathrm{m}}}} &
			\omega_{\mathrm{m}} - i(\kappa_2 + \beta) &
			-i\sqrt{\kappa_2 \kappa_3}\, e^{i\frac{2\pi d_{23}}{\lambda_{\mathrm{m}}}} \\[6pt]
			-i\sqrt{\kappa_1 \kappa_3}\, e^{i\frac{2\pi d_{13}}{\lambda_{\mathrm{m}}}} &
			-i\sqrt{\kappa_2 \kappa_3}\, e^{i\frac{2\pi d_{23}}{\lambda_{\mathrm{m}}}} &
			\omega_{\mathrm{m}} - i(\kappa_3 + \beta)
		\end{pmatrix}.
		\label{H_eff_3}
	\end{equation}
	Here we consider all magnon modes to be resonant at the same frequency $\omega_{\mathrm{m}}$, 
	with an identical intrinsic damping rate $\beta$. The spatial separation between the $i$th and $j$th magnon modes is denoted as $d_{ij}=|x_i-x_j|$, $\kappa_i~(i=1,2,3)$ represents the radiative decay rate of each magnon mode, and $\lambda_{\mathrm{m}}$ is the wavelength of the traveling photon mode at the magnon resonance frequency. 
	In the following, we assume that the magnon modes are equally spaced, i.e., $d_{ij}=d|i-j|$, 
	where $d$ is the distance between adjacent magnon modes. 
	The phase term $e^{i2\pi d|i-j|/\lambda_{\mathrm{m}}}$ directly influences 
	the coupling among the magnon modes and the reflection spectra of the system. As two specific spatial configurations, the Bragg and anti-Bragg arrays yield totally different reflection behaviors. The Bragg (anti-Bragg) condition is defined by the spatial separation $d$ between spin ensembles as
	\begin{equation}
		d = n\lambda_m/4,
		\label{d}
	\end{equation}
	where $n$ is an even (odd) integer. Substituting Eq.~\eqref{d} into Eq.~\eqref{vi}, the coupling vector $\mathbf{V}$ can be expressed as
	\begin{equation}
		\mathbf{V} = (\sqrt{\kappa_1}, \sqrt{\kappa_2}e^{i\frac{\pi n}{2}}, \sqrt{\kappa_3}e^{i\pi n}, \cdots, \sqrt{\kappa_N}e^{i\frac{\pi n(N-1)}{2}}).
		\label{v_bragg_and}
	\end{equation}
	When the magnon array satisfy the Bragg condition ($n$ is even), the relation
	\begin{equation}
		\mathbf{V}_{\text{Bragg}}^{\top} = \mathbf{V}_{\text{Bragg}}^{\dagger}
	\end{equation}
	holds, implying that left- and right-incident coupling vector are identical. Consequently, the reflection spectra are always symmetric under Bragg condition. This is because only the superradiance state can be detected in this system~\cite{Sheremet-23}. In our experiment, this property enables direct determination of relative phase between spin ensembles. For example, a $\pi$ phase difference between the first and third spin ensembles is verified by the symmetric reflection spectra shown in Figs.~3D and 3F of the main text. 
	
	In contrast, under the anti-Bragg condition (odd $n$), the relation $\mathbf{V}^{\top} = \mathbf{V}^{\dagger}$ no longer holds, resulting in asymmetric reflection. In our case, a $\pi/2$ phase difference between the first and second spin ensembles is confirmed by the asymmetric spectra shown in Figs.~3E and 3G of the main text. This is the key mechanism enabling the system to support unidirectional RL states. Under the anti-Bragg configuration, Eq.~\eqref{H_eff_3} reduces to
	\begin{equation}
		H_{\mathrm{eff}}^{\text{anti-Bragg}}=\left( \begin{matrix}
			\omega_{\mathrm{m}}-i(\kappa_1+\beta) & \sqrt{\kappa _1\kappa _2} & i\sqrt{\kappa _1\kappa _3} \\
			\sqrt{\kappa _1\kappa _2} & \omega_{\mathrm{m}}-i(\kappa_2+\beta) & \sqrt{\kappa _2\kappa _3} \\
			i\sqrt{\kappa _1\kappa _3} & \sqrt{\kappa _2\kappa _3} & \omega_{\mathrm{m}}-i(\kappa_3+\beta)
		\end{matrix} \right),
		\label{H_eff_anti}
	\end{equation}
	where coherent coupling occurs between nearest-neighbor magnon modes, while dissipative coupling arises between next-nearest neighbors. Using the mathematical formula introduced in Sec.~S5, we can obtain the RL Hamiltonian under the anti-Bragg condition. For right-incident input,
	\begin{equation}
		\mathbf{V}^{(r)}=(\sqrt{\kappa_1},\ i\sqrt{\kappa_2},\ -\sqrt{\kappa_3}), \qquad
		\mathbf{V}^{(r)}\big(\mathbf{V}^{(r)}\big)^{\top}=\kappa_1-\kappa_2+\kappa_3,
	\end{equation}
	and for left incidence,
	\begin{equation}
		\mathbf{V}^{(l)}=(\sqrt{\kappa_1},\ -i\sqrt{\kappa_2},\ -\sqrt{\kappa_3})
		=\big(\mathbf{V}^{(r)}\big)^{*}, \qquad
		\mathbf{V}^{(l)}\big(\mathbf{V}^{(l)}\big)^{\top}=\kappa_1-\kappa_2+\kappa_3.
	\end{equation}
	Following Eqs.~\eqref{w}–\eqref{S}, we define for each incidence direction
	\begin{equation}
		\mathbf{w}^{(r)}=
		\begin{pmatrix}
			\sqrt{\kappa_1} \\[6pt]
			i\sqrt{\kappa_2} \\[6pt]
			-\sqrt{\kappa_3}-\sqrt{\kappa_1-\kappa_2+\kappa_3}
		\end{pmatrix}, 
		\qquad
		\mathbf{w}^{(l)}=
		\begin{pmatrix}
			\sqrt{\kappa_1} \\[6pt]
			-i\sqrt{\kappa_2} \\[6pt]
			-\sqrt{\kappa_3}-\sqrt{\kappa_1-\kappa_2+\kappa_3}
		\end{pmatrix},
	\end{equation}
	and construct
	\begin{equation}
		\mathbf{S}^{r(l)}=\mathbf{I}_3-\frac{\mathbf{w}^{r(l)}\big(\mathbf{w}^{r(l)}\big)^{\top}}
		{\kappa_1-\kappa_2+\kappa_3+\sqrt{\kappa_3}\sqrt{\kappa_1-\kappa_2+\kappa_3}}.
	\end{equation}
	This satisfies $\big(\mathbf{S}^{r(l)}\big)^{\top}\mathbf{S}^{r(l)}=\mathbf{I}_3$ and 
	$\mathbf{S}^{r(l)}\big(\mathbf{V}^{r(l)}\big)^{\top}=\sqrt{\kappa_1-\kappa_2+\kappa_3}\,\mathbf{e}_3$.
	Taking the first two columns of $\big(\mathbf{S}^{r(l)}\big)^{\top}$ as $\mathbf{Q}^{r(l)}$, the reduced RL Hamiltonian is obtained as
	\begin{equation}
		H_{\mathrm{RL}}^{r(l)}=\big(\mathbf{Q}^{r(l)}\big)^{\top}H_{\mathrm{eff}}\,\mathbf{Q}^{r(l)}
		=\big[\omega_m-i\big(\beta\pm \tfrac{\Gamma_{R1}+\Gamma_{R2}}{2}\big)\big]\mathbf{I}+
		\begin{pmatrix}
			\pm i\tfrac{\Gamma_{R1}-\Gamma_{R2}}{2} & J_R\\[4pt]
			J_R & \mp i\tfrac{\Gamma_{R1}-\Gamma_{R2}}{2}
		\end{pmatrix}.
		\label{H_RZ_3}
	\end{equation}
	The matrix elements are given by
	\begin{equation}
		\Gamma_{R1}=\frac{2 \,\kappa_1\kappa_2\big[(\kappa_2-2\kappa_3)-2\sqrt{\kappa_3(\kappa_1-\kappa_2+\kappa_3)}\big]}{\big(\kappa_1-\kappa_2+\kappa_3+\sqrt{\kappa_3(\kappa_1-\kappa_2+\kappa_3)}\big)^2},
	\end{equation}
	\begin{equation}
		\Gamma_{R2}=-\frac{2 \,\kappa_2\big[\kappa_1(\kappa_1-\kappa_3)+\kappa_3(\kappa_2-2\kappa_3)-2\kappa_3\sqrt{\kappa_3(\kappa_1-\kappa_2+\kappa_3)}\big]}{\big(\kappa_1-\kappa_2+\kappa_3+\sqrt{\kappa_3(\kappa_1-\kappa_2+\kappa_3)}\big)^2},
	\end{equation}
	\begin{equation}
		J_R=-\frac{2\sqrt{\kappa_1\kappa_2}\,\big[(\kappa_1+\kappa_3)(\kappa_2-2\kappa_3)-(\kappa_1+2\kappa_3)\sqrt{\kappa_3(\kappa_1-\kappa_2+\kappa_3)}\big]}{\big(\kappa_1-\kappa_2+\kappa_3+\sqrt{\kappa_3(\kappa_1-\kappa_2+\kappa_3)}\big)^2}.
	\end{equation}
	The first term of Eq.~\eqref{H_RZ_3} represents a global frequency shift and uniform loss. The second term, being traceless with balanced gain and loss for the diagonal elements and symmetric coupling $J_R$ for the off-diagonal elements, constitutes a canonical parity-time (PT)-symmetric structure. This form makes the Hamiltonian $H_{\mathrm{RL}}^{r(l)}$ explicit and reveals the underlying mechanism for the emergence of exceptional points (EPs). The eigenfrequencies of $H_{\mathrm{RL}}^{r(l)}$ are given by
	\begin{equation}
		\omega_{\pm}^{r} = \omega_m - i\Big(\beta + \tfrac{\Gamma_{R1}+\Gamma_{R2}}{2}\Big) 
		\pm \sqrt{J_R^2 - \Big(\tfrac{\Gamma_{R1}-\Gamma_{R2}}{2}\Big)^2}
		\label{omega_RZ1}
	\end{equation}
	and
	\begin{equation}
		\omega_{\pm}^{l} = \omega_m - i\Big(\beta - \tfrac{\Gamma_{R1}+\Gamma_{R2}}{2}\Big) 
		\pm \sqrt{J_R^2 - \Big(\tfrac{\Gamma_{R1}-\Gamma_{R2}}{2}\Big)^2}.
		\label{omega_RZ2}
	\end{equation}
	Clearly, the eigenvalues in both directions share the same degeneracy condition
	\begin{equation}
		J_R=\frac{1}{2}|\Gamma_{R1}-\Gamma_{R2}|.
		\label{degeneracy_condition1}
	\end{equation}
	Substituting the explicit forms of $\Gamma_{R1}$, $\Gamma_{R2}$, and $J_R$ into Eqs.~\eqref{omega_RZ1} and~\eqref{omega_RZ2} yields the analytical dependence of the RL eigenfrequencies and reflection coefficient on the $\kappa_1$, $\kappa_2$, and $\kappa_3$. The resulting expressions are
	\begin{equation}
		R^{r(l)} = -i(\kappa_1-\kappa_2+\kappa_3)\frac{(\omega - \omega_+^{r,l})(\omega - \omega_-^{r,l})}{\det(\omega \mathbf{I} - H_{\mathrm{eff}}^{\text{anti-Bragg}})},
		\label{rforandback_new}
	\end{equation}
	where
	\begin{subequations}
		\begin{equation}
			\omega _{\pm}^{r}=\omega_m-i\beta+\frac{i\kappa _2\left( \kappa _1-\kappa _3 \right) \pm \sqrt{\kappa_2(4\kappa_1-\kappa_2)(\kappa _1+\kappa _3)\left(\kappa _3-\frac{\kappa _1\kappa _2}{4\kappa _1-\kappa _2}\right)}}{\kappa _1-\kappa _2+\kappa _3},
			\label{omega_RZ3}
		\end{equation}
		\begin{equation}
			\omega _{\pm}^{l}=\omega_m-i\beta+\frac{-i\kappa _2\left( \kappa _1-\kappa _3 \right) \pm \sqrt{\kappa_2(4\kappa_1-\kappa_2)(\kappa _1+\kappa _3)\left(\kappa _3-\frac{\kappa _1\kappa _2}{4\kappa _1-\kappa _2}\right)}}{\kappa _1-\kappa _2+\kappa _3}.
			\label{omega_RZ4}
		\end{equation}
	\end{subequations}
	When $\kappa_{1}>\kappa_{2}>\kappa_{3}$, equations~\eqref{omega_RZ3} and~\eqref{omega_RZ4} show that the degeneracy condition in both directions as 
	\begin{equation}
		\kappa_3 = \frac{\kappa_1\kappa_2}{4\kappa_1 - \kappa_2}.
		\label{degeneracy_condition2}
	\end{equation}
	The eigenvalues in two directions become different when $\kappa_1\neq\kappa_3$, leading to the asymmetric reflection spectra. Since the experimental probing frequency $\omega$ is purely real, only RL states with real eigenvalues ($\mathrm{Im}(\omega_{\pm}^{r(l)})=0$) can be observed in the spectra measurement.  However, the RL states in both direction cannot be observed simultaneously  since $\mathrm{Im}(\omega_{\pm}^{r}) \neq \mathrm{Im}(\omega_{\pm}^{l})$ when $\kappa_1 \neq \kappa_3$. Because $\kappa_1$ is much larger than $\kappa_3$ in our experiment, this is the fundamental reason we can observe the unidirectional RL state.  Thus, the emergence of the unidirectional RL EP corresponds to the pure real and dengerated RL eigenvalues in one direction. By substituting these two conditions into the RL Hamiltonians of two directions, we find that $H_{\mathrm{RL}}^{r}$ satisfies the PT symmetry, while the $H_{\mathrm{RL}}^{l}$ does not. This difference is the underlying mechanism enabling the construction of the unidirectional RL EP in our system. The above mathematical procedure yields the same solution as setting the numerator of the reflection coefficient to zero in the main text, but it clarifies the origin of this unidirectional second-order EP.

\subsubsection*{S7. The distinction between spectral in linear and decibel (dB) scales of reflection}
In Supplementary Fig.~\ref{Linear_dB}, we present the evolution of the reflection spectrum as $\kappa_3$ varies while $\kappa_1$ and $\kappa_2$ are kept fixed. Linear-scale representations are shown in Supplementary Figs.~\ref{Linear_dB}A and \ref{Linear_dB}AB, with their corresponding dB-scale counterparts provided in Figs.~\ref{Linear_dB}C and \ref{Linear_dB}D. At the unidirectional RL EP, the critical value is derived as $\kappa_3 = \frac{\kappa_1\kappa_2}{4\kappa_1 - \kappa_2}$.

\begin{figure*}[htp]
	\centering
	\includegraphics[scale=0.8]{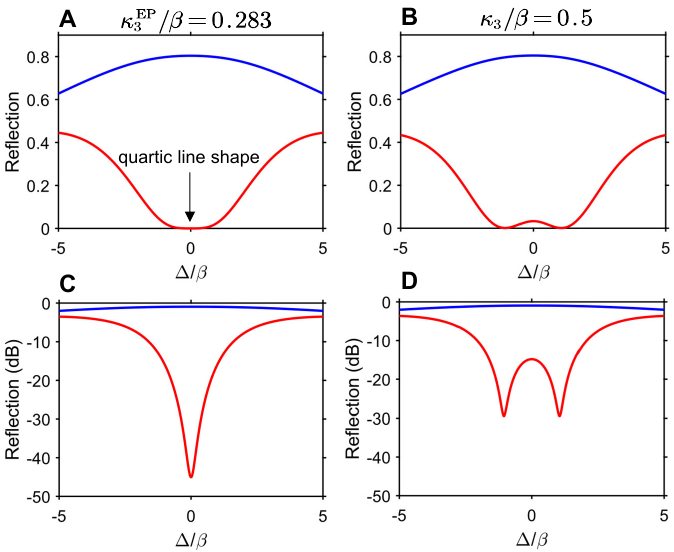}
	\caption{\textbf{Reflection spectra at and near the unidirectional RL EP.} (\textbf{A}) and (\textbf{C}) display the reflection spectra under RL EP conditions ($\kappa_3/\beta = 0.283$) on linear and dB scales, respectively. (\textbf{B}) and (\textbf{D}) show the spectra for $\kappa_3/\beta = 0.5$ on linear and dB scales, respectively. These results are acquired when $\kappa_1/\beta = 9$ and $\kappa_2/\beta = 1.1$. }
	\label{Linear_dB}
\end{figure*}

The characteristic quartic lineshape~\cite{Stone-19,Yang-21} is clearly resolved in Fig.~\ref{Linear_dB}A at the RL EP. The corresponding sharp dip on the dB scale in Fig.~\ref{Linear_dB}C further confirms the degeneracy of the reflectionless states. However, deviation from EP conditions eliminates the quartic dependence in Supplementary Fig.~\ref{Linear_dB}B, thereby its unidirectional bandwidth is significantly reduced.
\subsubsection*{S8. Cavity Magnonic System with Magnonic Mirrors}
\label{CMM}
Given that the first and third spin ensembles are separated by a distance of $\lambda_m/2$, the effective non-Hermitian Hamiltonian $H_{\text{eff}}^{\text{anti-Bragg}}$ can be transformed from the single-excitation basis $\{\sigma^{\dagger}_{i}|g_1g_2g_3\rangle\}$ to a new orthonormal basis $\{|B\rangle, |D\rangle, \sigma^{\dagger}_{2}|g_1g_2g_3\rangle\}$, where the bright and dark states are defined as
	\begin{subequations}
		\begin{equation}
			|B\rangle = \frac{\sqrt{\kappa_3}|egg\rangle - \sqrt{\kappa_1}|gge\rangle}{\sqrt{\kappa_1 + \kappa_3}},
		\end{equation}
		\begin{equation}
			|D\rangle = \frac{\sqrt{\kappa_3}|egg\rangle + \sqrt{\kappa_1}|gge\rangle}{\sqrt{\kappa_1 + \kappa_3}}.
		\end{equation}
	\end{subequations}	
	A unitary transformation matrix is applied as follows
	\begin{equation}
		U = \left(\begin{array}{ccc}
			\frac{\sqrt{\kappa_3}}{\sqrt{\kappa_1 + \kappa_3}} & 0 & \frac{\sqrt{\kappa_1}}{\sqrt{\kappa_1 + \kappa_3}} \\
			0 & 1 & 0 \\
			\frac{\sqrt{\kappa_1}}{\sqrt{\kappa_1 + \kappa_3}} & 0 & \frac{-\sqrt{\kappa_3}}{\sqrt{\kappa_1 + \kappa_3}}
		\end{array}\right),
	\end{equation}
	which transforms the effective Hamiltonian to
	\begin{equation}
		H_c = U^{\dagger} H_{\text{eff}}^{\text{anti-Bragg}} U = 
		\begin{pmatrix}
			-i\Gamma_D & J_D & 0 \\
			J_D & -i\gamma & J_B \\
			0 & J_B & -i\Gamma_B
		\end{pmatrix}.
	\end{equation}
	In this representation, the system is analogous to a cavity magnonic system, where the second magnon mode couples to two cavity modes defined by the bright and dark states, while the first and third spin ensemble, separated by $\lambda_m/2$, function as a pair of magnonic mirrors. Here, $J_{B(D)}$ are the coherent coupling strengths between the second magnon mode and the bright (dark) states, and the decay rates $\Gamma_{B(D)}$ correspond to their loss. The external decay of the second spin ensemble is denoted as $\gamma = \beta + \kappa_2$. These parameters are given by
	\begin{align*}
		J_D &= \frac{2\sqrt{\kappa_1\kappa_2\kappa_3}}{\sqrt{\kappa_1 + \kappa_3}}, &
		J_B &= \frac{\sqrt{\kappa_2}(\kappa_1 - \kappa_3)}{\sqrt{\kappa_1 + \kappa_3}}, \\
		\Gamma_D &= \beta, &
		\Gamma_B &= \beta + \kappa_1 + \kappa_3.
	\end{align*}

\subsubsection*{S9. Influence of the bright state on reflection spectra}

\begin{figure*}[t]
	\centering
	\includegraphics[scale=0.65]{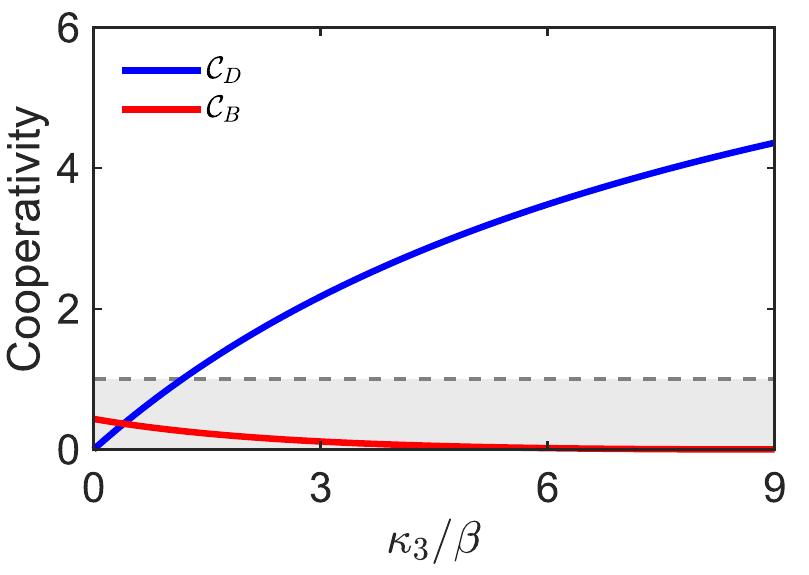}
	\caption{\textbf{Cooperativity of the cavity magnonic system.} Parameters: $\kappa_1/\beta = 9.1$, $\kappa_2/\beta = 0.93$. The gray region indicates $\mathcal{C} < 1$.}
	\label{cooperativity}
\end{figure*}
In the symmetric case with $\kappa_1 = \kappa_3$, the coupling strength $J_B = 0$, implying that the middle magnon mode only couples to the dark state \cite{Painter-19}. In this work, we focus on the asymmetric case with $\kappa_1 \gg \kappa_3$.	
	To quantify the effective coupling regime, we introduce the cooperativity parameters for dark and bright state coupling \cite{Sheremet-23}
	\begin{subequations}
		\begin{equation}
			\mathcal{C}_D = \frac{J_D^2}{\gamma \Gamma_D} = \frac{2\kappa_1\kappa_2\kappa_3}{\beta(\beta + \kappa_2)(\kappa_1 + \kappa_3)},
		\end{equation}
		\begin{equation}
			\mathcal{C}_B = \frac{J_B^2}{\gamma \Gamma_B} = \frac{\kappa_2(\kappa_1 - \kappa_3)^2}{(\beta + \kappa_2)(\beta + \kappa_1 + \kappa_3)(\kappa_1 + \kappa_3)}.
		\end{equation}
	\end{subequations}	
	It can be shown that $\mathcal{C}_B \leq 1$ for all non-negative values of $\kappa_1$, $\kappa_2$, $\kappa_3$, and $\beta$. Since $J_B$ is naturally smaller than $J_D$ and $\Gamma_B$ is much larger than $\Gamma_D$, $\mathcal{C}_B$ is almost always smaller than $\mathcal{C}_D$ over the range of $\kappa_3$ (Fig.~\ref{cooperativity}). Therefore, the dark state and middle magnon mode exhibit pronounced coherent behaviors, while the bright state behaves more like a dissipative reservoir rather than supporting coherent coupling. More importantly, $\mathcal{C}_D$ can be tuned to significantly exceed unity, allowing the strong coupling between the middle magnon mode and the dark state, as illustrated in Fig.~\ref{cooperativity}. To further assess the influence of asymmetry on our analysis, we plot the reflection coefficients for both propagation directions, along with the contrast ratio $C = |(|R^{r}|^2 - |R^{l})|^2/|(R^{r}|^2 + |R^{l}|^2)|$, versus the coupling strength \(\kappa_{3}\) and the detuning \(\Delta\) in Supplementary Fig.~\ref{RL_EP}. We observe that the bright state consistently manifests in both reflection spectra as a central, broad and high-reflectivity plateau that is insensitive to asymmetry between $\kappa_1$ and $\kappa_3$. This follows from its broad linewidth and strong coupling to the probe channel. Thus, the distinct RL dips observed for right incidence just signifies the coupling regimes transition of the equivalent magnonic mirror cavity system composed of the dark state and second magnon mode. 
\begin{figure*}[t]
	\centering
	\includegraphics[scale=1.1]{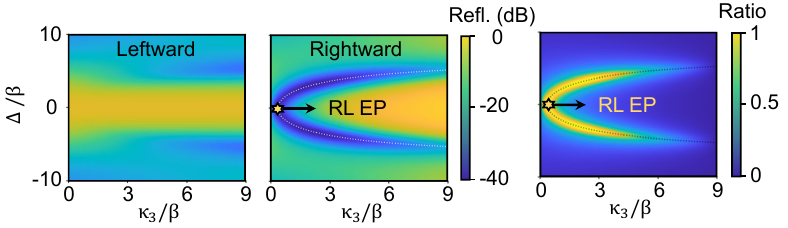}
	\caption{\textbf{Mapping of asymmetric reflection spectra.} Leftward reflection, rightward reflection, and their contrast ratio versus $\kappa_{\rm{3}}$. Dashed curves show the corresponding RL eigenfrequencies of the magnonic mirror array system. Parameters: $\kappa_1/\beta = 9$, $\kappa_2/\beta = 0.93$.}
	\label{RL_EP}
\end{figure*}
The above results illustrate that the bright state can be resonably neglected in the reflection spectra measurement. These unidirectional RL states provide a unique opportunity to directly observe magnonic mirror array (MMA) polariton behaviors via waveguide transmission in a single direction. This directionality persists until the system regains inversion symmetry, as illustrated in Supplementary Fig.~\ref{RL_EP} when $\kappa_{1}/\beta = \kappa_{3}/\beta \approx 9$.

\subsubsection*{S10. Fitting of the reflectionless state eigenvalues}
\begin{figure}
		\centering
		\includegraphics[width=0.8\textwidth]{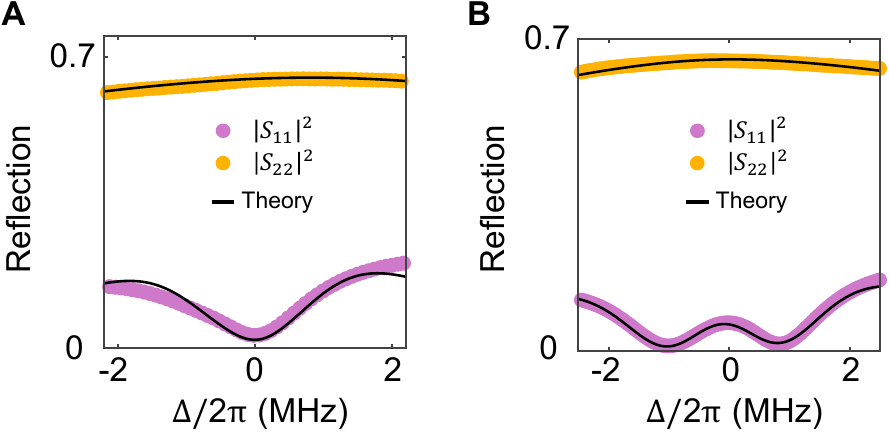}
		\caption{\textbf{Fitting of the RL-state eigenvalues.} (\textbf{A} and \textbf{B}) Representative reflection spectra measured from both waveguide ports (port 1 and port 2) at $\kappa_{3}/\beta = 0.173$ (A) and $\kappa_{3}/\beta = 0.73$ (B). The circular points indicate the experimental data, and the solid curves show the corresponding theoretical fits.}
		\label{fit}
	\end{figure}

Despite the quartic dip observed in the spectrum, the definitive verification of the RL EP relies on the degeneracy of the RL-state eigenvalues. These eigenvalues are extracted by fitting the reflection spectra measured under the condition that all three magnon modes are brought into resonance. To provide a clearer illustration of the fitting procedure, we additionally present two representative reflection spectra, distinct from those shown in Figs.~4G and 4I of the main text, corresponding respectively to the regimes before and far beyond the RL EP (Figs.~\ref{fit}A and \ref{fit}B). In the former case ($\kappa_{3}/\beta = 0.173$, Fig.~\ref{fit}A), the real parts of $\omega_{\rm{RL}}^{l,\pm}$ coalesce while the imaginary parts remain distinct. As a result, the $S_{11}$ spectrum features a single Lorentzian dip with near-zero reflection, rather than the characteristic quartic reflectionless dip that emerges at the RL EP. In the latter case ($\kappa_{3}/\beta = 0.73$, Fig.~\ref{fit}B), the real parts split more profoundly than in Fig.~4I, whereas the imaginary parts merge. Consequently, the $S_{11}$ spectrum displays two well-separated but identical Lorentzian reflection dips.
As $S_{22}$ remains a broad reflection peak throughout the variation of $\kappa_{3}$, the corresponding RL-state eigenvalues $\omega_{\rm{RL}}^{r,\pm}$ cannot capture the resonance characteristics of the spectrum~\cite{Han-23}. This behavior is fully consistent with our theoretical analysis in Supplementary Section~S9, which predicts that the RL states and the RL EP emerge unidirectionally in this system. The numerical fitting therefore provides further confirmation of the appearance of unidirectional RL EP.

\subsubsection*{S11. Device Photo}
\begin{figure}
	\centering
	\includegraphics[width=0.9\textwidth]{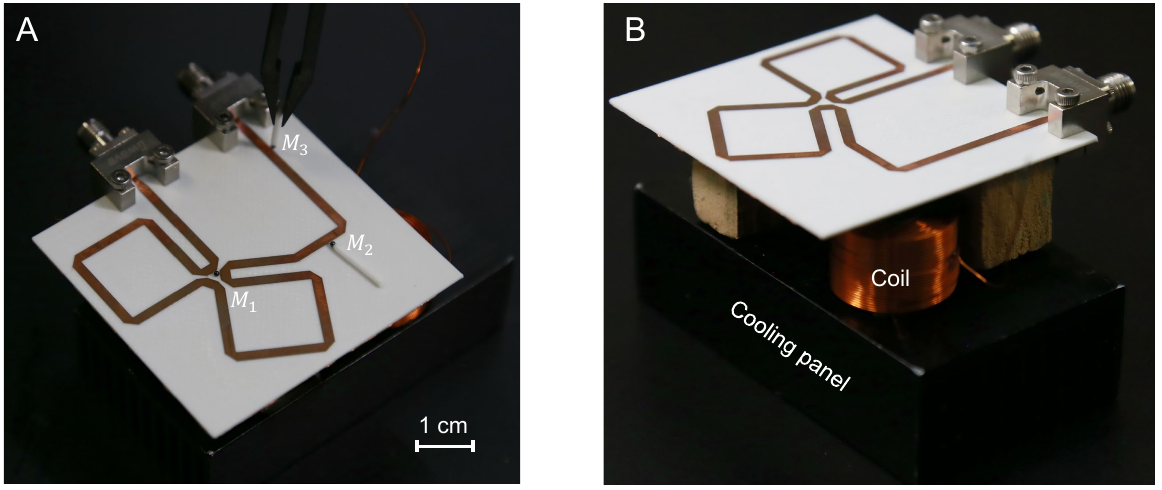}
	\caption{\textbf{Device photo.} (\textbf{A}) Top view photograph of the experimental setup, where the overall chip area is 540 mm × 520 mm. (\textbf{B}) Side view photograph of the experimental setup. The coil applies a local magnetic field at $M_2$ and a panel attached at the bottom is used to cool the coil.}
	\label{photo}
\end{figure}

Supplementary Fig. \ref{photo} shows a device photograph that clearly displays the magnonic mirror array used in our experiment.
\subsubsection*{S12. Phase-Dependent Reflection Asymmetry}

To investigate the influence of the phase on reflection asymmetry, we consider two magnon modes cooperatively coupled to the waveguide. As schematically illustrated in Supplementary Fig.~\ref{fig:phase_to_position}A, the relative phase of two YIG spheres is $\varphi = kd$, where $d = x_2 - x_1$ is the  separation. Under this configuration, the Hamiltonian $H_{\mathrm{eff}}$ and coupling vector $V$ take the following forms
	\begin{equation}
		H_{\mathrm{eff}}=
		\begin{pmatrix}
			\omega_m - i(\beta+\kappa_1) & -i\sqrt{\kappa_1\kappa_2}\, e^{i\varphi} \\
			-i\sqrt{\kappa_1\kappa_2}\, e^{i\varphi} & \omega_m - i(\beta+\kappa_2)
		\end{pmatrix}, 
		\qquad
		\mathbf{V} = 
		\begin{pmatrix}
			\sqrt{\kappa_1} \\
			\sqrt{\kappa_2}\, e^{i\varphi}
		\end{pmatrix}.
		\label{eq:Heff_V}
	\end{equation}
	By substituting \eqref{eq:Heff_V} into \eqref{rGeneral}, the rightward reflection coefficient can be deduced as
	\begin{equation}
		R^r = 
		\frac{\kappa_{2} e^{i2\varphi}\big[\beta-\kappa_{1}-i(\omega-\omega_m)\big] 
			+\kappa_1 \big[\beta+\kappa_{2}-i(\omega-\omega_m)\big]}
		{\kappa_{1}\kappa_{2} e^{i2\varphi} - 
			\big[\beta+\kappa_{1}-i(\omega-\omega_m)\big]
			\big[\beta+\kappa_{2}-i(\omega-\omega_m)\big]} .
		\label{eq:Rr}
	\end{equation}
	The corresponding leftward reflection coefficient $R^l$ follows by swapping the subscripts $1 \leftrightarrow 2$. Thus, the asymmetry of the reflection is defined as
	\begin{equation}
		A = |R^r|^2 - |R^l|^2 =\frac{8\beta\kappa_{1}\kappa_{2}\left(\kappa_{1}-\kappa_{2}\right)\sin ^{2}\varphi}{\left|\left(\beta-i(\omega-\omega_m)+\kappa_{1}\right)\left(\beta-i(\omega-\omega_m)+\kappa_{2}\right) e^{2i\varphi}-\kappa_{1}\kappa_{2}\right|^{2}},
	\end{equation}
	which directly reveals how the phase influences the degree of reflection asymmetry. As shown by the numerical results in Supplementary Fig.~\ref{fig:phase_to_position}B, the reflection spectra evolve continuously from asymmetric to symmetric as $\varphi$ varies from $\pi/2$ to $\pi$, confirming that the transition changes gradually rather than abruptly. 
\begin{figure*}[htb]
	\centering
	\includegraphics[width=0.9\textwidth]{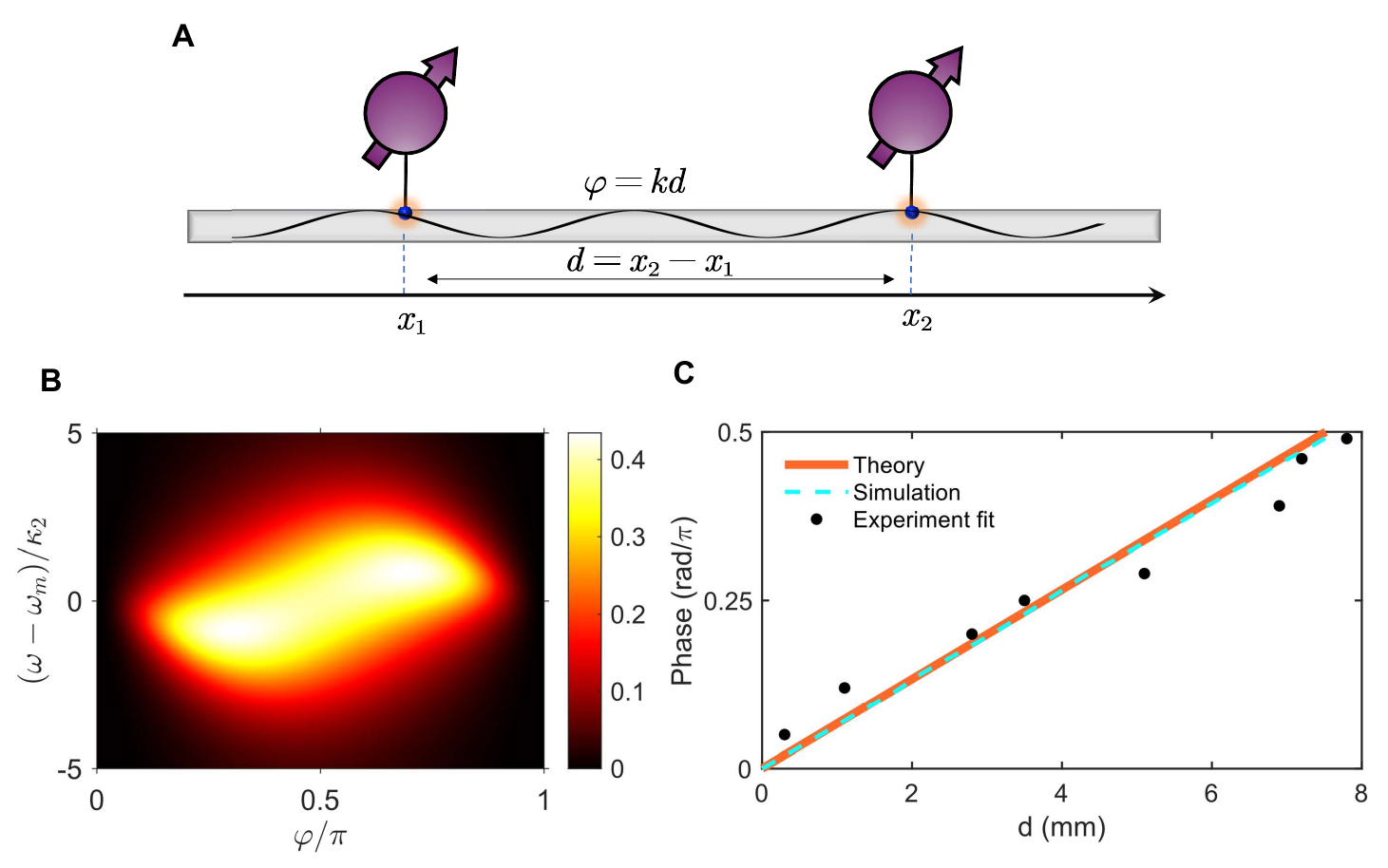}
	\caption{\textbf{Phase-dependent reflection asymmetry in the waveguide magnonic system.} 
			\textbf{(A)} Schematic of two YIG spheres coupled to the microstrip with separation $d$, introducing a phase $\varphi = k d$. 
			\textbf{(B)} Calculated reflection asymmetry $A = |R^r|^2 - |R^l|^2$ as a function of phase $\varphi$ and frequency detuning $(\omega-\omega_m)$. Parameters: $\kappa_1/\kappa_2=3,\,\beta/\kappa_2=1$. 
			\textbf{(C)} Experimentally extracted relation between position and phase. Dots, the solid and dashed lines represent experimental data, theoretical fit and simulation result, respectively.}
	\label{fig:phase_to_position}
\end{figure*}
Experimentally,  the phase was tuned by varying the distance between the two YIG spheres via step motors. By fitting the asymmetry between the measured rightward and leftward reflection spectra, we extracted the position-phase relation displayed as the dots in Supplementary Fig.~\ref{fig:phase_to_position}C. Since the effective wave number of the microstrip at the operating frequency $\omega/2\pi \approx 5.8$~GHz is constant ($k_{\mathrm{eff}} \approx 209$~m$^{-1}$), the position–phase dependence agrees well with a linear fit (solid line in Supplementary Fig.~\ref{fig:phase_to_position}C). This result is further validated by the CST simulations (dashed line in Supplementary Fig.~\ref{fig:phase_to_position}C).
\end{document}